\documentclass[useAMS,usenatbib]{mnras}

\usepackage{graphicx}
\usepackage{txfonts}

\title[Parameters of young stellar population]
      {Physical parameters of stellar population in star formation 
       regions of galaxies}

\author[A.~S.~Gusev et al.]
       {A.~S.~Gusev,$^{1}$\thanks{E-mail:gusev@sai.msu.ru}
        F.~Sakhibov,$^{2}$
        O.~V.~Egorov,$^{3,1}$
        V.~S.~Kostiuk,$^{4}$
        and E.~V.~Shimanovskaya$^{1}$ \\
 $^{1}$ Sternberg Astronomical Institute, Lomonosov Moscow State University, 
        Universitetsky pr. 13, 119234 Moscow, Russia \\
 $^{2}$ University of Applied Sciences, Technische Hochschule Mittelhessen, 
        61169 Friedberg, Germany \\
 $^{3}$ Astronomisches Rechen-Institut, Zentrum f\"{u}r Astronomie der 
        Universit\"{a}t Heidelberg, M\"{o}nchhofstra$\beta$e 12-14, 
        69120 Heidelberg, Germany \\
 $^{4}$ St.Petersburg State University, Universitetskaya nab. 7/9,
        St.Petersburg, 199034, Russia 
        }

\date{Accepted 2023 July 7. Received 2023 July 6; 
in original form 2022 December 21}

\begin{document}

\maketitle

\begin{abstract}
We present the results of a study of young unresolved stellar groupings 
(clusters, OB associations, and their complexes) associated with 
H\,{\sc ii}~regions, based on the coupling of spectroscopic, photometric 
and H$\alpha$ spectrophotometric observations of star formation regions. 
Along with our own observations, we use a part of the spectroscopic and 
H$\alpha$ data from the literature and open databases. The 
study is based on the catalogue of 1510 star formation regions with ages 
$\sim10-20$~Myr in 19 spiral galaxies, compiled by us earlier. We 
study the morphology of stellar groupings and their relation with the
associated H$\alpha$ emission region. Extinctions, gas chemical 
abundances, and sizes of star formation regions are measured. Using 
numerical SSP models computed for metallicities fixed from observations 
to intrinsic colours of the studied star formation regions, we
estimated ages and masses of stellar population of 400 young stellar 
groupings. Different relations between observational and physical 
parameters of the young stellar population in star formation regions are 
discussed.
\end{abstract}


\section{Introduction}

To understand processes of modern star formation in galaxies 
and to study the early evolution of star clusters, OB~associations 
and complexes, one needs to estimate physical and chemical parameters of 
young stellar groupings in star formation regions, including their age, 
mass, size, and metallicity. A galaxy star-forming region is a single 
mixture of newly formed star clusters, ionized gas, and clouds of molecular 
gas and dust. Star formation regions form a hierarchical structure on scales 
from a few to several hundreds of parsecs. The largest star formation 
regions are star complexes with typical sizes of about 300-700~pc 
\citep{elmegreen1996,efremov1998}. The diameters of the largest 
complexes reach 2~kpc \citep{elmegreen1996b}. These complexes are 
the largest coherent groupings of stars, clusters, and associations 
which are connected by the unity of the origin from the same 
H$_2$ supercloud \citep{efremov1989,efremov1995,elmegreen1994,
elmegreen2009,odekon2008,marcos2009}. On small scales, there are star 
clusters with sizes from a few parsecs which have been formed within 
dense cores of giant molecular clouds (GMCs). OB~associations and stellar 
aggregates with sizes from $\sim40$ to $\sim200$~pc occupy intermediate 
scales of star formation.

\begin{table*}
\caption[]{\label{table:sample}
The galaxy sample.
}
\begin{center}
\begin{tabular}{cccccccccccc} \hline \hline
Galaxy & Type & $B_t$ & $M(B)^a$ & $i$         & PA       & $R_{25}^b$ & 
$D$   & $A(B)_{\rm Gal}$ & $A(B)_{\rm in}$ & $n$ & Data$^c$ \\
       &      & (mag) & (mag)   & (degr)       & (degr)   & (arcmin)   & 
(Mpc) & (mag)        & (mag)       &   &        \\
1 & 2 & 3 & 4 & 5 & 6 & 7 & 8 & 9 & 10 & 11 & 12 \\
\hline
NGC~245   & SA(rs)b     & 12.72 & $-21.12$ & 21 & 145 & 0.62 & 53.8
& 0.097 & 0.04 &  33 & Ph$^1$ \\
NGC~266   & SB(rs)ab    & 12.27 & $-21.94$ & 15 &  95 & 1.55 & 63.8
& 0.252 & 0.01 &  19 & Ph$^1$+H$\alpha^2$ \\
NGC~628   & SA(s)c      &  9.70 & $-20.72$ &  7 &  25 & 5.23 &  7.2
& 0.254 & 0.04 & 497 & Ph$^3$+H$\alpha^4$+Sp$^{5-15}$ \\
NGC~783   & Sc          & 13.18 & $-22.01$ & 43 &  57 & 0.71 & 70.5
& 0.222 & 0.45 &  32 & Ph$^{16,17}$+Sp$^{12}$ \\
NGC~2336  & SAB(r)bc    & 11.19 & $-22.14$ & 55 & 175 & 2.51 & 32.2
& 0.120 & 0.41 &  48 & Ph$^{18}$+H$\alpha^{19}$+Sp$^{12}$ \\
NGC~3184  & SAB(rs)cd   & 10.31 & $-19.98$ & 14 & 117 & 3.79 & 10.2
& 0.060 & 0.02 & 109 & Ph$^{20,21}$+H$\alpha^{21}$+Sp$^{5,8,13,22}$ \\
NGC~3726  & SAB(r)c     & 10.31 & $-20.72$ & 49 &  16 & 2.62 & 14.3
& 0.060 & 0.30 & 129 & Ph$^{23,24}$+H$\alpha^{25}$+Sp$^{22}$ \\
NGC~4136  & SAB(r)c     & 11.92 & $-18.38$ & 22 &  30 & 1.20 &  8.0
& 0.066 & 0.05 &  11 & Ph$^{26}$+Sp$^{22}$ \\
NGC~5351  & SA(r)b      & 12.57 & $-21.16$ & 60 & 101 & 1.20 & 51.1
& 0.074 & 0.40 &  16 & Ph$^{24,27}$+Sp$^{22}$ \\
NGC~5585  & SAB(s)d     & 10.94 & $-18.73$ & 53 &  34 & 2.13 &  5.7
& 0.057 & 0.38 &  69 & Ph$^{28}$+H$\alpha^{29}$+Sp$^{22}$ \\
NGC~5605  & (R)SAB(rs)c & 12.58 & $-20.86$ & 36 &  65 & 0.81 & 44.8 
& 0.318 & 0.15 &   6 & Ph$^{30}$ \\
NGC~5665  & SAB(rs)c    & 12.25 & $-20.42$ & 53 & 151 & 0.95 & 31.1
& 0.091 & 0.35 &   7 & Ph$^{30}$ \\
NGC~6217  & (R)SB(rs)bc & 11.89 & $-20.45$ & 33 & 162 & 1.15 & 20.6 
& 0.158 & 0.22 &  28 & Ph$^{1,31}$+H$\alpha^2$+Sp$^{12}$ \\
NGC~6946  & SAB(rs)cd   &  9.75 & $-20.68$ & 31 &  62 & 7.74 &  5.9
& 1.241 & 0.04 & 320 & Ph$^1$+H$\alpha^{32}$+Sp$^{5-7,11,33,34}$ \\
NGC~7331  & SA(s)b      & 10.20 & $-21.68$ & 75 & 169 & 4.89 & 14.1
& 0.331 & 0.61 & 100 & Ph$^{1,35}$+H$\alpha^{32}$+Sp$^{9,12}$ \\
NGC~7678  & SAB(rs)c    & 12.50 & $-21.55$ & 44 &  21 & 1.04 & 47.8
& 0.178 & 0.23 &  22 & Ph$^{1,36}$+Sp$^{12}$ \\
NGC~7721  & SA(s)c      & 11.11 & $-21.18$ & 81 &  16 & 1.51 & 26.3
& 0.121 & 0.98 &  36 & Ph$^1$ \\
IC~1525   & SBb         & 12.51 & $-21.89$ & 48 &  27 & 0.97 & 69.6 
& 0.410 & 0.24 &  15 & Ph$^{37}$ \\
UGC~11973 & SAB(s)bc    & 13.34 & $-22.47$ & 81 &  39 & 1.73 & 58.8
& 0.748 & 0.85 &  13 & Ph$^1$+Sp$^{38}$ \\
\hline
\end{tabular}\\
\end{center}
\begin{flushleft}
$^a$ The absolute magnitude of a galaxy corrected for Galactic extinction 
and inclination effects. \\
$^b$ The radius of a galaxy at the isophotal level 25 mag\,arcsec$^{-2}$ in 
the $B$ band corrected for Galactic extinction and inclination effects. \\ 
$^c$ References: 1 -- \citet{gusev2015},2 -- \citet*{epinat2008}, 
3 -- \citet{bruevich2007}, 4 -- \citet{gusev2013b}, \\
5 -- \citet*{mccall1985}, 6 -- \citet{belley1992}, 7 -- 
\citet*{ferguson1998}, 8 -- \citet{zee1998}, \\
9 -- \citet*{bresolin1999}, 10 -- \citet{rosales2011}, 11 -- 
\citet{cedres2012}, 12 -- \citet{gusev2012}, 13 -- \citet{sanchez2012}, \\ 
14 -- \citet{berg2013}, 15 -- \citet{berg2015}, 16 -- \citet{gusev2006a}, 
17 -- \citet{gusev2006b}, 18 --  \citet{gusev2003}, 19 -- 
\citet{young1996}, \\
20 -- \citet{larsen1999b}, 21 -- \citet{gusev2002b}, 22 -- SDSS, 23 -- 
\citet{gusev2002}, 24 -- \citet{gusev2018}, 25 -- \citet{knapen2004}, \\
26 -- \citet*{gusev2003b}, 27 -- \citet{gusev2004}, 28 -- 
\citet*{bruevich2010}, 29 -- \citet{dale2009}, \\
30 -- \citet*{artamonov2000}, 31 -- \citet{artamonov1999}, 32 -- 
\citet{gusev2016}, 33 -- \citet{garsia2010}, 34 -- \citet{gusev2013}, \\
35 -- \citet{regan2004}, 36 -- \citet*{artamonov1997}, 37 -- 
\citet*{bruevich2011}, 38 -- \citet*{gusev2020}.
\end{flushleft}
\end{table*}

This paper focuses on studying the stellar groupings in star 
formation regions of rather distant galaxies (see 
Table~\ref{table:sample}). The angular resolution of our observations 
$\sim1-1.5$~arcsec corresponds to the linear resolution 30-40~pc for 
the nearest galaxies NGC~628, NGC~5585, and NGC~6946, and 350-400~pc 
in the faraway galaxies NGC~783 and IC~1525. It does not allow 
us to separate the young star clusters and OB~associations even in the 
nearest galaxies: smaller star clusters are observed as star-like 
objects with diameters of 30-40~pc. In more distant galaxies, we can 
observe star formation regions with sizes of 200--300~pc and larger, 
i.e. star complexes. Star clusters, embedded in star formation regions, 
are dense aggregates of young stars, formed at essentially the same time 
in the same region of space \citep*{zwart2010}. In our previous paper 
\citep{gusev2016}, we found that the minimal masses of the studied star 
clusters in the nearest galaxies NGC~628 and NGC~6946 are 
$\approx1\cdot10^4 M_{\odot}$. According to \citet{zwart2010}, star 
clusters that are more massive than $\sim10^4 M_{\odot}$ are determined 
as young massive clusters. \citet{gieles2011} showed that the youngest 
(age $\le10$~Myr) clusters and associations are poorly separated. Thus, 
most of the objects studied here are young massive clusters (associations) 
or complexes of young star clusters. Hereinafter, we will call
the studied stellar populations in star formation regions the 'stellar 
groupings'. It shall be understood that this common term encompasses 
different types of young objects, from giant complexes of clusters and 
stars to OB~associations and star clusters.

\begin{figure*}
\vspace{1.7mm}
\resizebox{1.00\hsize}{!}{\includegraphics[angle=000]{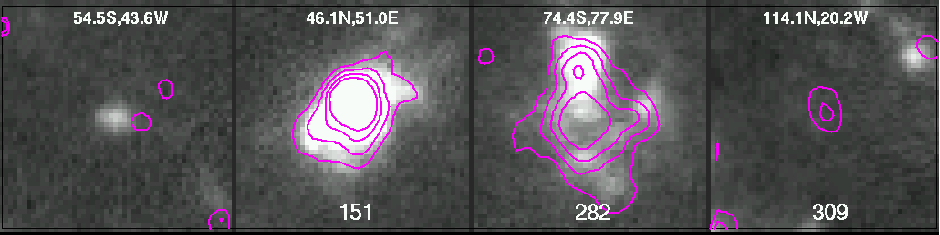}}
\caption{H$\alpha$ images of star formation regions in NGC~628 with 
superimposed isophotes 22.0, 21.5, 21.0, and 20.5~mag\,arcsec$^{-2}$ in 
the $U$ band are shown. Galactocentric coordinates of centres of the 
images (top) and serial numbers from the catalogue, if any, (bottom) 
are given. The size of the images is $13.6\times13.6$~arcsec$^2$. 
North is upwards and east is to the left. See the text for details.
}
\label{figure:map1}
\end{figure*}

A star formation region goes through several stages of evolution during 
first tens Myr of its life, from the stage when young stars are completely 
obscured by their dusty gas cocoons to the stage of a young star cluster 
with no evidence of the ionized gas \citep{lada2003}. 
\citet{whitmore2011} developed an evolutionary classification scheme of 
star clusters based on {\it Hubble Space Telescope (HST)} observations of 
M83. Star clusters become visible in optical bands since the age of 
$\sim2.5$~Myr \citep{Kim2021,Kim2023}. The authors showed that in clusters 
with ages between 2 and 4~Myr, the ionized gas is observed in the same place 
as the cluster stars. Clusters of ages $\approx4-5$~Myr are surrounded 
with small H\,{\sc ii}~bubbles whose radii are equal to 7-20~pc 
\citep{whitmore2011}. The phase of the partially embedded cluster blowing 
a bubble of gas is rather short, it lasts for about 1-3~Myr 
\citep{hollyhead2015,Kim2021,Kim2023}. Star clusters with ages of 
$> 5$~Myr are surrounded by a large ionized gas bubble.
The radii of the bubbles are larger than 20~pc. The ionized gas is 
not detected around star clusters of ages $> 10$~Myr. 
Figure~\ref{figure:map1} illustrates this evolutionary sequence on the 
sample of our young unresolved objects.

A study of the earliest stages of star clusters, OB~associations and their 
complexes and estimation of physical parameters therein are difficult tasks 
because of the impact of gas and dust on the observations. Perhaps the most 
difficult task is to estimate ages of stellar populations. Usually, 
2D or 3D spectroscopic or photometric data, or their combination are used 
for estimating ages of unresolved extragalactic star clusters. The 
spectroscopic method involves both estimation of spectral age indicators 
(e.g., equivalent widths EW(H$\alpha$) and EW(H$\beta$), 
[O\,{\sc iii}]/H$\beta$ ratio, He\,{\sc ii} emission lines, etc.) and a 
direct comparison of spectra with synthetic spectra of different ages 
\citep*{copetti1986,bastian2005,bastian2006,bastian2009,
konstantopoulos2009,wofford2011}. The photometric method involves
comparison of multicolour photometry data for clusters with predictions 
of evolutionary synthesis models \citep*{searle1980,elson1985,
bresolin1996,chandar2010,hollyhead2015,hollyhead2016,
adamo2017,turner2021}.

Comparison shows that ages, evaluated for the same star clusters using data of
spectral and photometric observations are in a fairly good agreement 
\citep[see, e.g.,][]{searle1980,whitmore2011,wofford2011}. However, 
\citet{kim2012} who studied resolved stellar populations in star 
formation regions of M83 found that correlation between the ages of star 
clusters determined from individual stars in the region and the ages 
obtained via integrated colours using a standard photometric method is not 
very strong. This discrepancy, in addition to the reasons indicated 
by \citet{kim2012} (H$\alpha$ emission impact, selection effects for 
stars, overlay of isochrones for 1 and 3~Myr), we tend to explain also 
by the use of a continuously populated IMF and the presence of a small 
but significant range of ages of stars in the clusters.

Spectroscopic techniques usually provide age estimates 
\citep[see, e.g.,][]{garsia2010}, however the method allows determination 
of ages for a limited number of objects. One of the main challenges 
in photometric age estimation is accounting for the effect of gas 
and dust on observations. A lack of independent data on the chemical 
abundance and extinction in the clusters leads to 'age--extinction' and 
'age--metallicity' degeneracies in the comparative analysis with the 
theoretical evolutionary models of star clusters \citep{scalo1986}. Moreover, 
continuum and line emissions from the ionized gas are strong enough 
to affect the integrated broad-band photometry \citep{reines2010}.

Age and mass estimates based on long-slit spectroscopic observations are 
correct only if the radiation of stars, which form the continuum of the 
spectrum, spatially coincides with the ionized gas emission. This situation 
is observed in star formation regions younger than $\approx5$~Myr 
\citep{whitmore2010}. The combination of optical photometric, H$\alpha$ 
spectrophotometric and spectroscopic observations provides us with the 
necessary data to separate spatially the radiation from gas and 
stellar components. This makes it possible to take into account 
the contribution of the gas to the optical photometric bands and to find 
objects in which the light extinction for the stars is equal to the light 
extinction for the emission of ionized gas \citep{gusev2016,gusev2018}.

Note that for stellar clusters (simple stellar population (SSP) systems) 
with masses less than $5\cdot10^3-10^4 M_\odot$, stochastic effects in 
the discrete random population IMF start playing a key role and applying 
the 'standard' mode of continuous IMF models will not be correct 
\citep{whitmore2010,piskunov2011}. According to \citet{cervino2013}, IMF 
discreteness significantly affects the luminosity and colour of the 
cluster. The strength of this effect depends on the wavelength and is 
particularly strong in IR wavelengths, where the effect of discreteness 
is noticeable up to $M_{\rm cl} \sim 10^7 M_{\odot}$ masses, rather 
than $10^4 M_{\odot}$ as at optical wavelengths in the $V$ band. 
Thus, the estimation of physical parameters of stellar groupings, 
using photometric methods with a continuously populated IMF is correct 
only for massive star clusters.

This paper presents the conclusive part of our project of comprehensive 
study of star formation regions in the selected 19 spiral galaxies. The 
results of our own spectroscopic observations of 103 H\,{\sc ii}~regions 
in eight galaxies were presented in the previous papers 
\citep*{gusev2012,gusev2013,gusev2020}.
In \citet{gusev2016} we estimated physical parameters of stellar 
population in H\,{\sc ii}~regions using a combination of spectroscopic 
\citep{gusev2012,gusev2013} and photometric observations (see 
Table~\ref{table:sample}). We derived properties of extracted emission 
spectra of H\,{\sc ii}~regions and estimated their extinctions, chemical 
abundances, and the relative contributions of nebular continuum and 
emission lines to the total observed flux. These data were used to 
obtain the luminosities and colour indices of stellar groupings, corrected 
for extinction and nebular emission contribution, i.e. 'true' (intrinsic) 
colours and luminosities of stellar population. As a result, we were able to 
estimate ages and masses for $\approx60\%$ of clusters (complexes) of our 
sample. Extinctions for $\approx35\%$ of objects were overestimated. This is 
due to the fact that the key assumption of our method, the equality of the 
light extinction for stars, $A({\rm stars})$, and the light extinction for 
ionized gas, $A({\rm gas})$, is not satisfied for a significant number of 
young clusters. 

The fact that the extinction in a gaseous medium is up to 2 times 
higher than the stellar one is a well-known fact for a long time 
\citep{calzetti2001}. \citet{sakhibov1990} empirically investigated the 
discrepancy between the extinction of gas emission and the extinction of 
stellar light in the giant H\,{\sc ii}~regions, i.e. in star formation 
complexes in galaxies M33, LMC and NGC~2403. They found that, in most cases, 
$A({\rm Balmer})\equiv A({\rm gas})$ is higher than $A({\rm stars})$. 
Such a result can be explained in terms of the quite uneven distribution 
of obscuring material \citep{caplan1986}. So, using correction of the 
observed colours of stars in the star formation complexes for extinction 
in the Balmer lines can result in a bias in the colours of the stars in 
the complexes toward the blue part of the spectrum, thus distorting the 
parameters of star formation derived from these colours. 
Objects with $A({\rm gas}) \ne A({\rm stars})$ have a high 
nebular emission contribution in the $U$, $B$, $V$ bands ($>40\%$) and an 
extremely high ${\rm EW(H}\alpha)>1500$\AA. Visually, in these regions a 
spatial displacement between the photometric centres of the stars in the 
broad bands and of the gas emissions in H$\alpha$ line is observed 
(see the third image from left in Fig.~\ref{figure:map1}).

Later, we presented the catalogue of 1510 young stellar groupings 
associated with H\,{\sc ii}~regions in 19 galaxies in \citet{gusev2018} 
using multicolour photometric and H$\alpha$ (H$\alpha$+[N\,{\sc ii}]) 
spectrophotometric observations. This catalogue is available in electronic 
form\footnote{\url{http://lnfm1.sai.msu.ru/\~gusev/sfr_cat.html}}. In 
the same paper we modified our extinction and age estimation techniques using 
H$\alpha$ morphology as an additional indicator. This method was developed 
in \citet{whitmore2011}.

The goal of this study is to estimate the physical parameters, such as 
mass and age, of the stellar population in the star formation regions of 
the galaxies of our sample, using additional spectral data for 
H\,{\sc ii}~regions taken from the literature and open databases.

Most of the stellar groupings studied in this paper have an age of 
$\sim1-10$~Myr, i.e. they are objects with H$\alpha$ emission, visible 
at optical wavelengths. In addition, we studied young star clusters with 
colour indices typical for stellar populations younger than 10~Myr without 
visible H$\alpha$ emission including the cases for which H$\alpha$ data are 
absent for the galaxy. These young objects may be older than 10~Myr 
(see Section~\ref{sect:phys} for details).

The sample of selected galaxies is based on our $UBVRI$ photometric survey
of 26 galaxies \citep{gusev2015}. Numerous star formation regions are 
observed visually in 19 of them. The sample is presented in 
Table~\ref{table:sample}, where data on the Galactic extinction, 
$A(B)_{\rm Gal}$, are taken from the 
NED\footnote{\url{http://ned.ipac.caltech.edu/}} database, and the other 
parameters are taken from the 
LEDA\footnote{\url{http://leda.univ-lyon1.fr/}} database \citep{paturel2003}.
The morphological type of the galaxy is listed in column (2). The apparent
and absolute $B$ magnitudes are presented in columns (3) and (4). The
inclination and position angles are given in columns (5) and (6). The
isophotal radii in units of arcmin are shown in column (7). The 
adopted distances are given in column (8). The Galactic extinction 
and the dust extinction due to the inclination of a galaxy are listed 
in columns (9) and (10). The number of identified star formation regions 
in the galaxy is shown in column (11). A presence of photometric (Ph) and 
H$\alpha$ spectrophotometric observations of the galaxies, as well as 
spectrophotometric and spectroscopic (Sp) data for the star formation 
regions and the references to them are given in column (12). The adopted 
value of the Hubble constant in the study is equal to 
$H_0 = 75$ km\,s$^{-1}$Mpc$^{-1}$.

\section{Data and methods used}
\label{sect:data}

The algorithm and techniques for data reduction, criteria for selecting 
star formations regions, and evolutionary synthesis models used were 
described in detail in our previous paper 
\citep{gusev2013,gusev2016,gusev2018,gusev2019}. 
In this paper, we describe only the new data and models used, as well as 
the data considered earlier very briefly.

\subsection{Observational data}

Most part of our own observations was published earlier (see 
Table~\ref{table:sample}). Additionally, we used FITS images of the 
galaxies which were taken from the NED database, as well as spectroscopic 
data from the Sloan Digital Sky Survey 
DR13\footnote{\url{http://www.sdss.org/dr13/}} \citep{albareti2017} and 
from the literature (see references in Table~\ref{table:sample}).

\subsubsection{Photometric and spectrophotometric H$\alpha$ images}

Earlier we carried out photometric observations of 19 galaxies, studied 
here, and published the analysis of the photometric data  
\citep[see][and references therein, see also notes in 
Table~\ref{table:sample}]{gusev2015,gusev2018}.

Spectrophotometric H$\alpha$ observations of NGC~3184 and 
H$\alpha$+[N\,{\sc ii}] observations of NGC~628, NGC~6946, and NGC~7331 
were described in our previous papers 
\citep{gusev2002b,gusev2013b,gusev2016}. FITS images, obtained with 
narrow-band interference H$\alpha$+[N\,{\sc ii}] or H$\alpha$ filters 
for another five galaxies from our sample, were found in the NED database 
(see references in Table~\ref{table:sample}).

We used the H$\alpha$+[N\,{\sc ii}] FITS image of NGC~3726 obtained by 
\citet{knapen2004}. Parameters for absolute calibration of 
H$\alpha$+[N\,{\sc ii}] flux to units of erg\,s$^{-1}$cm$^{-2}$ were found 
in descriptors of the FITS file. Absolute calibration of the FITS image of 
NGC~5585 from \citet{dale2009} was done according to the data in 
descriptors of the FITS file. Additionally, we checked the calibration using 
integrated H$\alpha$+[N\,{\sc ii}] fluxes of NGC~5585 measured in 
\citet{james2004} and \citet{kennicutt2008}. For a study of 
H\,{\sc ii}~parameters in NGC~2336, we used the H$\alpha$+[N\,{\sc ii}] 
FITS image obtained in \citet{young1996}. FITS file descriptors and results 
of integrated H$\alpha$+[N\,{\sc ii}] photometry from \citet{young1996} 
were used for absolute flux calibration. For two galaxies (NGC~266 and 
NGC~6217) we used H$\alpha$ FITS images published in \citet{epinat2008}. 
Their absolute calibrations were carried out using the results of 
integrated H$\alpha$ photometry of \citet{epinat2008} for NGC~266 and 
NGC~6217, and integrated spectrophotometry of \citet{james2004} for 
NGC~6217. Note that the absolute calibration uncertainty of~NGC 266 and 
NGC~6217 can reach $\approx20-25\%$. This accuracy, however, is sufficient 
for estimates of nebular emission contributions in total fluxes from 
star formation regions in galaxies \citep{gusev2016,gusev2018}.

\subsubsection{Spectroscopic data for star formation regions}
\label{sect:spectra}

The results of spectroscopic observations of 103 H\,{\sc ii}~regions in 
eight galaxies have already been published in our previous papers 
\citep{gusev2012,gusev2013,gusev2020}. An explicit description of the 
observational  data reduction is given in \citet{gusev2012}.
Additionally, in this paper we used data of emission-line 
spectrophotometry, integral field spectroscopy, and long-slit spectroscopy 
for H\,{\sc ii}~regions in the galaxies in our sample from the literature 
and SDSS (see notes in Table~\ref{table:sample}).

Among 19 galaxies of our sample, we found spectral data for 
H\,{\sc ii}~regions in NGC~628, NGC~3184, NGC~6946, and NGC~7331 in the 
literature. Some H\,{\sc ii}~regions in NGC~3184, NGC~3726, NGC~4136, 
NGC~5351, and NGC~5585 were observed in the SDSS project.

At the first stage, we cross-identified the H\,{\sc ii}~regions observed 
by different authors. Then these regions were identified 
with the objects from our catalogue (see Sections~\ref{sect:select}, 
\ref{sect:catalog}).

A description of calculation of the extinction coefficient, $c$(H$\beta$), 
from the measured Balmer decrement was presented in \citet{gusev2012}.

Oxygen abundances, O/H, in H\,{\sc ii}~regions were obtained using the 
reddening-corrected fluxes in the main emission lines 
[O\,{\sc ii}]$\lambda$3727+3729, [O\,{\sc iii}]$\lambda$4959+5007, 
[N\,{\sc ii}]$\lambda$6548+6584, and [S\,{\sc ii}]$\lambda$6717+6731. 
Because different sets of emission lines were measured in 
different studies, we used four different empirical calibration methods. 
In order of priority, these are S-calibration \citep{pilyugin2016}, 
R-calibration \citep{pilyugin2016}, O3N2 calibration 
\citep{pettini2004,marino2013}, and H\,{\sc ii}-ChiMistry method 
\citep{perez2014}. We also took our O/H data from \citet{gusev2012} 
measured using NS-calibration \citep{pilyugin2011}.

For objects, observed in several studies, we took weighted 
averages of the measured abundances, extinctions, and H$\alpha$ 
equivalent widths with weights inversely proportional to their relative 
measurement uncertainties.

We used all available data (our, SDSS, and from the literature) to find 
the mean values, with one exception. Spectrophotometric measurements of 
\citet{belley1992} for NGC~628 and NGC~6946 were used only for objects
which were not observed by any other authors. This is because
the spectrophotometry results of \citet{belley1992} 
are not as accurate as spectroscopic ones for individual objects
\citep[see][for details]{bruevich2007}. We also do not include 
recently published measurements made for H~\textsc{ii} regions in NGC~628 
within inner 1.5 effective radii based on IFU spectroscopy with MUSE/VLT 
\citep{Groves2023}, but compare the results in Section~\ref{sect:phys}.

\subsection{Sample selection}
\label{sect:select}

The procedure of selection of young stellar groupings was described 
in detail in \citet{gusev2018}. We note briefly that the preliminary 
selection of bright star formation sources from $B$ and H$\alpha$ 
images of galaxies was carried out with the use of the 
{\sc SExtractor}\footnote{\url{http://sextractor.sourceforge.net/}} 
program. We searched stellar groupings associated with H\,{\sc ii}~regions 
and young star clusters with colour indices corresponding to stellar 
populations younger than 10~Myr.

The final selection criteria for the objects, included in our catalogue, 
have been explained in \citet{gusev2018}. The selected young stellar 
groupings must satisfy one of the following conditions: (i) those, which 
form close pairs with the nearest H\,{\sc ii}~regions: the angular 
separation between photometric centres of the stars in $B$ band and 
the gas emission in H$\alpha$ is less than 1.5~arcsec (in 9 galaxies with 
obtained H$\alpha$+[N\,{\sc ii}] or H$\alpha$ images), (ii) those, for which 
the emission spectra are measured (in 13 galaxies with obtained spectroscopic 
data), (iii) those, which have corrected for the Galactic extinction and 
inclination effects $(U-B)_0^i<-0.537$~mag (in 16 galaxies for which
$U$ images have been obtained), (iv) those, which have $(B-V)_0^i<-0.043$~mag 
(in NGC~4136, NGC~5605, NGC~5665, and outer part of NGC~7331).

Remark that the ambiguity of age from $U-B$ and $B-V$ values exists for 
stellar systems with ages between 6 and 40~Myr. Stellar groupings with 
$(U-B)_0^i$ colour indexes ranging from $-0.75$ to $-0.5$ can be both older 
and younger than 10~Myr \citep{gusev2018}.

\subsection{Young stellar groupings and gas-to-stars morphology}

As we noted in \citet{gusev2016}, the technique, that we use to determine 
the age and mass of the stellar component of star formation regions, has 
some limitations. The spatial displacement between photometric centres 
of stars and of gas emissions in star formation regions leads to 
incorrectly estimated extinction and overestimated contribution of nebular 
emission in optical broad-bands. Physical parameters (age and mass) 
of stellar population in star formation regions can be correctly retrieved
only provided that the optical radiation from stars spatially coincides 
with ionized gas emission. A typical sample of such regions is shown on 
the second image from left in Fig.~\ref{figure:map1}.

If both optical broad-band and H$\alpha$ images of a galaxy are available, 
it is possible to detect a presence or an absence of the spatial displacement 
by direct comparison of the positions of photometric centres in these bands. 
For galaxies, that were not observed in H$\alpha$ line, we 
can suspect the displacement by spectral features, such as an extremely 
large ${\rm EW(H}\alpha)$ ($>1500$\AA), an extremely high nebular emission 
contribution ($>40\%$) in the shortwave optical bands, an extremely large 
Balmer decrement, giving unrealistically 'blue' colour indices 
\citep{gusev2016}.

For objects with a star-like profile, we accept that the photometric 
emission centers of stars and gas coincide if the distance between them 
in a plane does not exceed 0.5~arcsec \citep[see][for details]{gusev2018}.

In addition to the objects where the optical emission from stars 
coincides with the H$\alpha$-emission from ionized gas, we can obtain 
physical parameters of young star clusters without a visible H$\alpha$ 
emission. These star clusters have the extinction that is close to 
zero \citep{whitmore2011}, thus we can assume 
$A({\rm stars})=A_{\rm Gal}+A_{\rm in}$, where $A_{\rm Gal}$ is the 
Galactic extinction and $A_{\rm in}$ is the dust extinction due to the 
inclination of a galaxy (see columns (9) and (10) in 
Table~\ref{table:sample}). A sample of such regions is shown on the right 
image in Fig.~\ref{figure:map1}.

We have classified the young stellar groupings studied here as follows: \\
class~2 -- optical radiation from stars coincides with ionized gas 
emission (second image from left in Fig.~\ref{figure:map1}); \\
class~1 -- photometric (stellar) radiation centre is displaced from gas 
emission centre (third image from left in Fig.~\ref{figure:map1}); \\
class~0 -- no gas emission within the area of optical radiation from 
stars (forth image from left in Fig.~\ref{figure:map1}); \\
class --1 -- no H$\alpha$ data.

\subsection{Comparison with synthetic models}

We described in detail the algorithm for correction of observational 
photometric fluxes for contribution from nebular continuum and emission 
lines in \citet{gusev2016}. To briefly summarize: we determined the 
relative contributions of the stellar and nebular continua, and gas 
emission lines to the total observed flux in the $UBVRI$ bands following 
\citet{sakhibov1990}.

We used the emission line ratios for every star formation region in 
our sample (see Section~\ref{sect:spectra}) to derive 
electron temperatures and metallicities in the H\,{\sc ii} regions. The 
fluxes for the non-measured emission lines were calculated based on the 
derived estimations of the emission measures, using the equations given 
in \citet{kaplan1979} and \citet{osterbrock1989}. A total of 18 main 
emission lines were taken into account. The contribution from the gas 
line emission was computed through the summation of the emission line 
intensities in a given photometric band.

The relative contribution of the nebular continuum was estimated using 
the equations for the continuum emission near the limits of the hydrogen 
series emission, two-photon and free-free emissions, given in 
\citet{lang1978,kaplan1979,brown1970,osterbrock1989}.

We used spectrophotometric H$\alpha$ (H$\alpha$+[N\,{\sc ii}]) fluxes 
(see Section~~\ref{sect:catalog}) for the absolute calibration of the 
emission line spectroscopic fluxes. For two galaxies without H$\alpha$ 
photometry, NGC~4136 and NGC~5351, we multiplied the absolute fluxes, 
obtained within the SDSS aperture, by a factor, calculated as the ratio 
of the flux in the $R$ band within the aperture that we used for every 
H\,{\sc ii}~region to the flux within the area of the SDSS aperture. 
The reliability of this procedure was discussed in \citet{gusev2016}.

Obtained 'true' photometric parameters of the star groups, i.e. colours 
and magnitudes corrected for the extinction and gas contribution, were 
compared with SSP evolutionary sequences using Salpeter IMF with 
a mass range from $0.15$ to $100 M_{\odot}$.

\begin{figure*}
\vspace{9mm}
\resizebox{1.00\hsize}{!}{\includegraphics[angle=000]{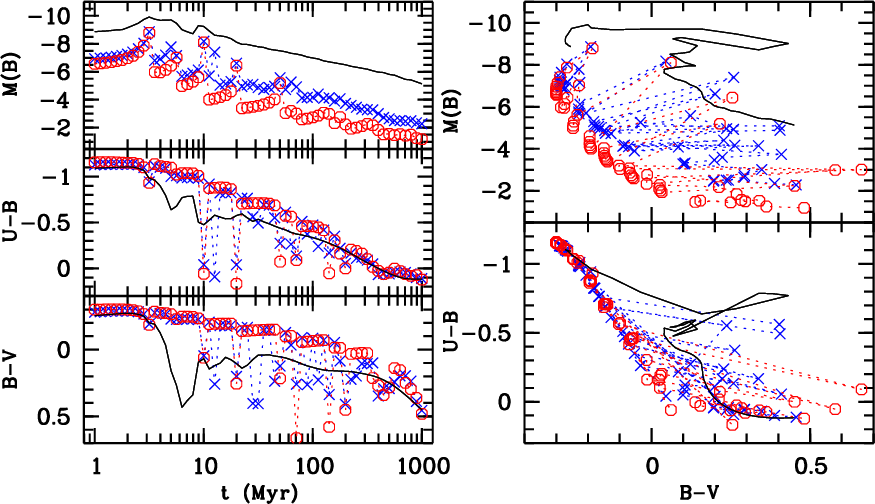}}
\caption{Samples of evolutionary sequences of the {\it Standard} SSP 
mode (continuously populated IMF) for a stellar system with $Z=0.012$ and
$M=1\cdot 10^4 M_{\odot}$ (black curves), and the 
{\it Extended} SSP mode (randomly populated IMF) for stellar systems 
with $Z=0.012$ and $M=1\cdot 10^3 M_{\odot}$ (blue crosses connected by 
dotted lines) and $M=500 M_{\odot}$ (red circles connected by dotted lines). 
Variations of absolute magnitude $M(B)$ and colour indices $U-B$ and $B-V$ 
versus age (left) and colour-magnitude and colour-colour diagrams (right) 
for the synthetic evolutionary sequences are shown. See the text for 
details.
}
\label{figure:evolt}
\end{figure*}

We used a database of stellar evolutionary tracks and isochrones 
provided by the Padova group 
\citep{bertelli1994,girardi2000,marigogirardi2007,marigo2008} via the 
online server CMD\footnote{\url{http://stev.oapd.inaf.it/cgi-bin/cmd/}}. 
We used the sets of stellar evolutionary tracks from version~2.8 
\citep{bressan2012,tang2014,chen2014,chen2015}.

For young massive clusters with $M>1\cdot10^4 M_{\odot}$, models in 
{\it Standard} modes have been developed adopting the technique described 
in \citet{piskunov2011}. The standard mode reproduces properties of standard 
SSP models with a continuously populated IMF, while the extended mode allows 
to take into account the influence of a randomly populated IMF.

As we discussed earlier in \citet*{gusev2014}, the multiple structure of 
unresolved star complexes does not affect their integrated colour indices 
and therefore the estimations of the age of a stellar 
population.

Multicolour photometry provides a useful tool for constraining  
masses and ages of stellar populations in star formation regions. Here 
we use the method of the minimisation of 'observed minus computed' 
$(O-C)$ parameters
\begin{eqnarray}
O-C=[[(U-B)_{\rm obs}-(U-B)_{\rm model}]^2+ \nonumber \\
+[(B-V)_{\rm obs}-(B-V)_{\rm model}]^2
+[M(B)_{\rm obs}-M(B)_{\rm model}]^2]^{1/2},
\label{equation:criterion}
\end{eqnarray}
described in \citep{gusev2007,gusev2016}, where under 
the concept of 'observed parameters' we place the 'true' colours $U-B$ 
and $B-V$, and $B$ luminosities. We did not use the $V-R$ and $V-I$ 
colour indices because, in the case of star formation regions, the $R$ 
and $I$ fluxes are weakly sensitive to changes in age, and actual 
observational errors lead to large uncertainties. 
Moreover, the stochastic effects of the stellar luminosity function 
are noticeable up to $M_{\rm cl} \sim 10^7 M_{\odot}$ masses, rather 
than $10^4 M_{\odot}$ as at optical wavelengths in $U$, $B$, and $V$ 
bands \citep{cervino2013}. The exception is a few objects without gas 
emission within the area of optical radiation from stars, for which 
observations in the $U$ band are unavailable. For them, we used the 
'true' colours $B-V$ and $V-R$.

The stellar population models, computed for $Z$, independently obtained 
from observations, are presented in the form of a grid of models for a 
broad range of variation of parameters $t(i)$ and $M(j)$, where the 
indices $i$, $j$ are the numbers of rows and columns in a two dimensional 
grid of physical parameters. The table step $h$ of the 
$\log t$ parameter variation is 0.05~dex. The initial table 
step of the $\log M$ for the first iteration depends on the range of 
luminosity variations of star formation regions in a given galaxy:
\begin{eqnarray}
h_{\log M}=(\log M_{\rm max}-\log M_{\rm min})/N, \nonumber
\end{eqnarray}
where $N$ is the number of the evolutionary sequences 
simulated for $N$ values of cluster masses within a given mass interval. 
For every node $(i,j)$, the value of the $(O-C)_{i,j}$ parameter was 
calculated. The second step is the search for the grid node in which the 
$(O-C)_{i,j}$ parameter has a minimum value.

Note that the value of the parameter $(O-C)_{i,j}$ 
corresponds to the distance of the investigated stellar cluster from the 
grid node $(i,j)$ in the three-dimensional photometric space of 
($U-B$, $B-V$, $M_B$). The minimum value of $(O-C)_{i,j}$ corresponds 
to the distance between the object under study and the nearest node in 
the photometric value space. This lowest parameter value 
$(O-C)^{\rm min}_{i,j}$ had to be less than the errors of the observed 
colours and luminosities. Otherwise, the next iteration was carried out, 
in which the mass interval was halved and centred according to the 
results of the previous iteration. Thus, beginning with the second 
iteration, for each stellar grouping under study, a particular grid of 
models was simulated according to the input observational data 
($M(B)_{\rm obs}$, $(U-B)_{\rm obs}$, $(B-V)_{\rm obs}$, $Z$). With 
each successive iteration, the density of the model grid was increased. 
The number of iterations per object needed to achieve the requirement 
$(O-C)^{\rm min}_{i,j}$ less than the observational error of photometric 
quantities, $\Delta_{U-B}$, $\Delta_{B-V}$, $\Delta_{M_B}$, ranged from 2 
to 4. The values of age $t(i)$ and mass $M(j)$ corresponding to the 
selected node $(i,j)$ with the minimum value of parameter 
$(O-C)^{\rm min}_{i,j}$, were taken as the age $t_{\rm cl}$ and mass 
$M_{\rm cl}$ for the stellar grouping under study. Simultaneous constraint 
using Eq.~\ref{equation:criterion} and true colours and luminosities in 
the $U$, $B$, $V$ bands, which are most sensitive to changes of age and 
mass, with a grid of models simulated for metallicity, which is fixed 
from independent observations, helps to avoid ambiguities associated with 
degenerations of 'metallicity-age', 'absorption-age', 'luminosity-mass' 
and ambiguity in the estimates of physical properties, age $t_{\rm cl}$ 
and mass $M_{\rm cl}$, within the adopted model.

The model grid of {\it Extended} mode was constructed 
using Monte Carlo simulations by which random variations of the discrete 
IMF depending on a given model star cluster mass were generated. For 
every given cluster mass, a discrete IMF was generated using a 
pseudo-random number generator. Note that with this random sampling of 
the discretely populated IMF for a fixed mass value of the stellar 
grouping, the number of stars, $N_{\rm stars}$ in that grouping is also 
fixed. Then, using this randomly chosen discrete 
IMF, we have calculated an evolutionary sequence of 68 models of 
{\it Extended} mode for every given cluster mass with a $\log t$-step of 
0.05 in the interval $\log t = 5.9 - 9.3$ and metallicity $Z$ fixed from 
the observations. For each calculation of a randomly sampled discrete IMF, 
a random seed was used, also obtained using a pseudo-random number 
generator. When comparing the observed colours and luminosity of a given 
object with that of a model, each iteration uses $N=50$ evolutionary 
sequences of 'discrete' models of {\it Extended}  mode. The number of 
iterations per object ranged from 2 to 4. So the number of simulations of 
a randomly sampled discrete IMF per object varies from 50 to 200. 
Number of simulated {\it Extended} mode models for each pair of mass 
$M_{\rm cl}$ and age $t_{\rm cl}$ estimates ranged from 6800 to 13600.

The errors of the age and mass estimates for the case of 
continuous IMF models have been calculated as follows. Using the 
evolutionary sequences for the star cluster colour indices in the $U$, $B$, 
$V$ bands, simulated for a fixed model grid node $(i,j)$, the coefficients 
of the third- or fourth-degree interpolation polynom were calculated for 
a time interval corresponding to the selected node $(i,j)$ with a minimum 
value of the parameter $(O-C)_{i,j}$. Knowing the functional (polynomial) 
correlation between colour and age, and the observational error of the 
colours used, we applied the Gauss law of error propagation to determine 
the accuracy of the age estimates. Similarly, using the functional 
correlation between the model luminosity and the cluster mass, as well as 
the observational errors of the integrated luminosities, the accuracies 
of the mass estimates were defined.

The IMF discreteness significantly affects the luminosity 
and colours of the cluster, as manifested by flashes and fluctuations in 
the evolutionary path of the cluster's photometric parameters, caused by 
the appearance of red giants. There is also a systematic bias between 
luminosities and colours of main sequence clusters and the predictions by 
standard SSP models 
\citep[see Fig.~\ref{figure:evolt} and more details in][]{piskunov2011}. 
The luminosity evolution curve of the discrete cluster model has the form of 
tilted oscillations and consists of relatively short time scale intervals 
of recurrent events. During a single time interval, there are a slow, 
gradual increase in cluster luminosity and an almost instantaneous outburst, 
caused by the evolution of the brightest star in the main sequence and 
its eventual transformation into a bright, short-lived red supergiant. 
After the supergiant's decline, the process is repeated by the evolution 
of the next brightest star on the main sequence and its transformation 
into a red giant. Note that the behaviour of the colour and luminosity 
evolution curves of the discrete model, described above, is stronger in the 
case of small cluster masses $M_{\rm cl}$, when the number of red giants 
is rare and the clusters spend most of the time as clusters with stars of 
the main sequence. As the cluster mass $M_{\rm cl}$ increases, the colour 
and luminosity evolution curves of the discrete model converge to those of 
the standard continuous model.

The errors of the age and mass estimates for the case 
of discrete IMF models have been calculated as follows. Based on the model 
colours corresponding to the selected node $(i,j)$ with the minimum value 
of parameter $(O-C)^{\rm min}_{i,j}$, the corresponding interval on the 
colour evolution curve of the discrete model between two 'red flashes' 
was selected and the coefficients of the interpolation polynomial were 
calculated. Then, as in the case of the continuous model, knowing the 
functional relationship between age and colour, as well as colour 
observation errors, we applied the Gauss law of error propagation to 
determine the accuracy of the age estimates. Similarly, using the 
functional correlation between the model luminosity and the cluster mass 
within the interval between two 'red flashes', as well as observational 
errors of the integrated luminosities, the accuracies of the mass 
estimates were determined.

In summary, the errors of age and mass estimates, 
calculated here, take into account the influence of colour and luminosity 
observation errors only. The influence of the accuracy of the choice of a
model grid node, on the estimates of ages and masses, we have not 
considered, just making sure that the minimum parameter 
$(O-C)^{\rm min}_{i,j}$ must not exceed the observational error of the 
colours and luminosities, $\Delta_{U-B}$, $\Delta_{B-V}$, $\Delta_{M_B}$. 
We have therefore not detected the effect of increasing inaccuracy when 
considering discrete models of small masses.

\begin{figure}
\vspace{5mm}
\resizebox{1.00\hsize}{!}{\includegraphics[angle=000]{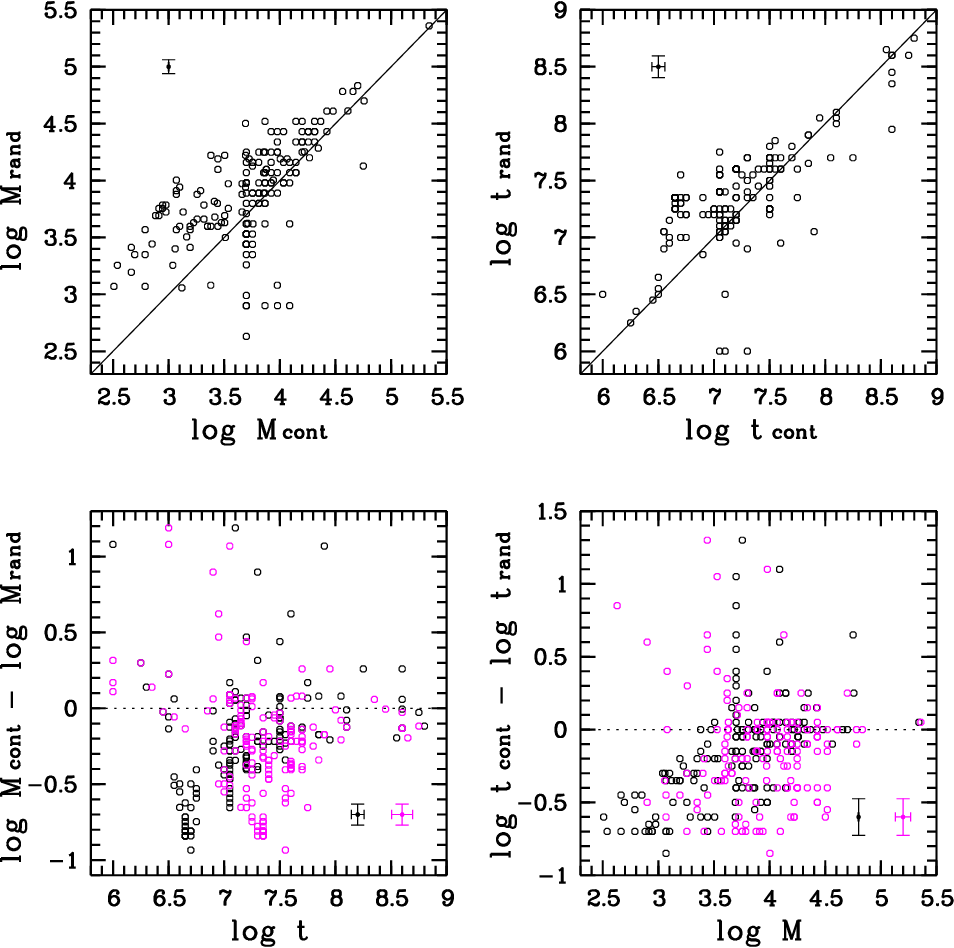}}
\caption{Comparison between masses (top-left panel) and ages (top-right 
panel) retrieved using evolutionary sequences with continuously and 
randomly populated IMF. One-to-one lines are shown. Dependencies 'difference 
between masses estimated using continuously and randomly populated IMF, 
$\log M_{\rm cont}-\log M_{\rm rand}$, over age, $\log t$' (bottom-left 
panel), and 'difference between ages estimated using continuously and 
randomly populated IMF, $\log t_{\rm cont}-\log t_{\rm rand}$, over mass, 
$\log M$' (bottom-right panel). Objects with ages (masses), estimated 
using continuously populated IMF, are shown by black circles, and ones 
with ages (masses), estimated using randomly populated IMF, are 
indicated by magenta circles in the bottom panels. Mean error bars 
are given.
}
\label{figure:imf}
\end{figure}

We compared the mass and age estimates, obtained using the continuously 
and randomly populated IMF, in Fig.~\ref{figure:imf}. The top diagrams 
show that $M_{\rm rand}$ and $t_{\rm rand}$ are systematically larger 
than $M_{\rm cont}$ and $t_{\rm cont}$, respectively. This difference 
decreases for high-massive ($M>5\cdot10^4 M_{\odot}$) and ageing 
($t>50$~Myr) stellar groupings. The differences between 
the mass estimates, obtained from continuous and random models for groups 
in the mass range of $5\cdot10^3 - 10^4 M_{\odot}$, are larger than for 
groups of higher mass. Above we have already noted the effect of IMF 
discreteness on the luminosity of the cluster, which is manifested by a 
systematic bias between luminosities of main sequence clusters with 
randomly populated IMF and the predictions by standard SSP models noted 
in \citet{bruzual2002} and discussed thoroughly in \citet{piskunov2011} 
and \citet{cervino2013}. The strength of this bias between the 
luminosities of the discrete and standard models is stronger at low 
masses and decreases with increasing cluster mass, which is apparent in our 
estimates of the masses of the stellar groupings explored here.

The maximum differences in age estimates are observed for objects with 
$t=5-12$~Myr (Fig.~\ref{figure:imf}). This difference 
results from the fact that in this age interval, the number of red giants 
in the case of the randomly populated IMF is rare and most of the cluster 
stars are main sequence stars. Models with a continuously populated IMF 
and ages $t=5-12$~Myr always contain red giants, which shift their colours 
towards red relative to those of models with a discrete IMF. With 
increasing age, the colour bias between continuous and discrete IMF 
models decreases. The bias also decreases at low ages $t<5$~Myr 
\citep[see fig.~7 in][]{piskunov2011}. The use of various methods for 
determining ages in the range of $5-10$~Myr gives results that differ 
by an order of magnitude or more 
\citep*[see, e.g.,][]{kim2012,popescu2012,messa2018}.

The maximum difference in $\log M_{\rm cont}-\log M_{\rm rand}$ is 
observed for objects with $t_{\rm cont}=5-12$~Myr and 
$t_{\rm rand}=10-30$~Myr (Fig.~\ref{figure:imf}). For the youngest 
($t<4$~Myr) and oldest ($t>50$~Myr) stellar groupings, it usually does 
not exceed 0.2~dex. The age difference 
$\log t_{\rm cont}-\log t_{\rm rand}$ is systematically 
negative for low-massive objects with $M<3\cdot10^3 M_{\odot}$. For most 
stellar groupings with $M>5\cdot10^3 M_{\odot}$, this difference does not 
exceed 0.2~dex (Fig.~\ref{figure:imf}).

In general, the data obtained in Fig.~\ref{figure:imf} correspond to 
the conclusions of \citet{whitmore2010} and \citet{piskunov2011} 
about the need to use models with a randomly populated IMF for star 
clusters with $M<1\cdot10^4 M_{\odot}$, where the IMF 
discreteness effect is particularly strong. At low masses,  models with 
a discrete IMF remain relatively long time like a main sequence cluster 
model, free of red giants, due to their small number. Continuous models 
cannot resemble main sequence clusters because they always contain a 
fraction of red giants. Therefore, at small masses, the luminosity of a 
branch of the main sequence cluster can be $2-3$~mag below the luminosity 
of a continuous track and colours are correspondingly bluer, 
$\Delta(B-V)\approx0.1-0.5$~mag \citep{piskunov2011}.

As noted by \citet{cervino2013}, the mean value of 
model simulations with a randomly sampled discrete populated IMF 
converges to the results of a 'standard' model with a continuously 
populated IMF. It is noted, however, that the 'random sampling' mode 
gives the entire distribution of possible age and mass values compared 
to the 'standard mode' for a set of photometric parameters fixed from 
the observations. At larger masses, on the other hand, the associated 
distributions approach Gaussians and the relative variance decreases 
\citep[see][and references therein]{cervino2013}, so only the mean (and 
variance) is required for inference. Indeed, as the star cluster mass 
increases, the IMF population density increases, converging to a 
continuously populated IMF density and the results converge to the 
inferences of the 'standard' model. Thus, when at 
$M_{\rm cl} \ge 10^4 M_{\odot}$ the bias between the discrete and 
continuous models is comparable or less than the errors of our 
observations and the application of continuously populated IMF models 
is more rational to reduce computational time.

In closing Section~\ref{sect:data}, we note again, that 
the bias between the discrete and continuous models is consistent with the 
above noted systematic excess of $M_{\rm rand}$ over $M_{\rm cont}$, 
especially at low masses, where the number of red flashes in the case of 
a discrete IMF is rare and the luminosity of the cluster is determined by 
the emission of main sequence stars. Since the continuous IMF models have 
for a given mass a luminosity excess compared to discrete mass models, the 
former option requires lower luminosity in order to agree both models. 
Older age estimates, obtained with the discrete model, can be similarly 
explained by the blue colour bias of the discrete models with respect to 
continuous option.

\section{Results}

\subsection{Catalogue of young stellar groupings}
\label{sect:catalog}

The catalogue is available in electronic form at 
http://lnfm1.sai.msu.ru/$\sim$gusev/sfr\_cat.html  and is also 
presented as the additional online material on the paper page of the 
MNRAS website.

The following data are presented in the catalogue columns: \\
(1) ID of the region; \\
(2) galaxy name (NGC, IC, or UGC); \\
(3) ID of the object within a galaxy; \\
(4,~5) apparent coordinates in the plane of the sky, with respect to the 
galaxy centre, in units of arcseconds; positive values correspond to the 
northern (4) and western (5) positions; \\
(6,~7) deprojected galactocentric distances in units of kpc (6) and in 
units of isophotal radius $R_{25}$ (7), where $R_{25}$ is the radius at 
the isophotal level 25 mag\,arcsec$^{-2}$ in the $B$ band corrected 
for the Galactic extinction and inclination effects; \\
(8) apparent total $B$ magnitude; \\
(9) absolute magnitude $M(B)$, $M(B) = B - 5\log D - 25$, where $D$ is an 
adopted distance in units of Mpc (see Table~\ref{table:sample}); \\
(10) $B$ magnitude uncertainty; \\
(11--18) apparent colour indices $U-B$ (11), $B-V$ (13), $V-R$ (15), and 
$V-I$ (17) with their uncertainties (12, 14, 16, 18); \\
(19,~20) logarithm of spectrophotometric H$\alpha$+[N\,{\sc ii}] flux 
(19), where the flux is in units of erg\,s$^{-1}$cm$^{-2}$, for all 
galaxies except NGC~266, NGC~3184, and NGC~6217, and logarithm of H$\alpha$ 
flux for NGC~266, NGC~3184, and NGC~6217, and their uncertainties (20); \\
(21) absolute magnitude $M(B)_0^i$, corrected for the Galactic extinction 
and inclination effects; \\
(22--25) the colour indices $(U-B)_0^i$ (22), $(B-V)_0^i$ (23), 
$(V-R)_0^i$ (24), and $(V-I)_0^i$ (25), corrected for the Galactic 
extinction and inclination effects; \\
(26) the same as column (19), but corrected for the 
Galactic extinction and inclination effects,
$F$(H$\alpha$+[N\,{\sc ii}])$_0^i$ ($F$(H$\alpha$)$_0^i$); \\
(27) $R-$H$\alpha$ index, 
$R-{\rm H}\alpha = R+2.5\log F({\rm H}\alpha$+[N\,{\sc ii}]) for all 
galaxies except NGC~266, NGC~3184, and NGC~6217, and 
$R-{\rm H}\alpha = R+2.5\log 1.35F({\rm H}\alpha$) for NGC~266, 
NGC~3184, and NGC~6217, where $R$ is in magnitudes, and 
$F$(H$\alpha$+[N\,{\sc ii}]), $F$(H$\alpha$) are the fluxes in units 
of erg\,s$^{-1}$cm$^{-2}$; \\
(28) gas-to-stars morphology (2 -- optical radiation from stars coincides 
with ionized gas emission (class~2), 1 -- photometric (stellar) radiation 
centre is displaced from the centre of gas emission (class~1), 0 -- no gas 
emission within the area of optical radiation from stars (class~0), --1 -- 
no H$\alpha$ data); \\
(29,~30) extinction $A(B)$ and its uncertainty $\Delta A(B)$ in units 
of magnitude calculated from Balmer decrement; \\
(31,~32) equivalent width EW(H$\alpha$) in units of \AA\, and 
its uncertainty; \\
(33,~34) logarithm of equivalent width EW(H$\alpha$) and its 
uncertainty; \\
(35,~36) metallicity $Z$ and its uncertainty; \\
(37) relative contribution of nebular continuum and emission lines 
to the total observed flux in $B$ band, 
$I_B({\rm gas})/I_B({\rm total})$; \\
(38,~39) 'true' absolute magnitude $M(B)_{\rm true}$, corrected for 
extinction and nebular emission contribution, and its uncertainty; \\
(40--47) 'true' colour indices $(U-B)_{\rm true}$ (40), 
$(B-V)_{\rm true}$ (42), $(V-R)_{\rm true}$ (44), and 
$(V-I)_{\rm true}$ (46), corrected for extinction and nebular emission 
contribution, and their uncertainties (41, 43, 45, 47); \\
(48,~49) the same as columns (19, 20), but corrected for extinction 
$A$, $I$(H$\alpha$+[N\,{\sc ii}]) ($I$(H$\alpha$)), and 
their uncertainties; \\
(50,~51) age $t$ in units of Myr, and its uncertainty; \\
(52,~53) mass $M$ in units of $10^4 M_{\odot}$, and its uncertainty; \\
(54) estimated diameter in units of pc; \\
(55) structure of the region (1 -- separate object with a star-like profile, 
2 -- double object, 3 -- triple object, 4 -- separate object with a diffuse 
profile, 5 -- ring structure, 6 -- complex structure (more than three 
separate objects), 10...60 -- the same as 1...6, but the object is a 
brighter part (core) of a more extended star forming region).

We give the parameters of H$\alpha$+[N\,{\sc ii}] (H$\alpha$) lines: 
$F$, $F_0^i$, $I$ fluxes, and equivalent widths EW(H$\alpha$) for all 
H\,{\sc ii}~regions associated with stellar groupings, including cases 
where (i) the photometric radiation centre is displaced from the gas emission 
centre (class~1) and (ii) the gas emission is absent within the area of 
radiation from stars (class~0). At the same time, $R-$H$\alpha$ index was 
not calculated for the class~0 objects.

Gas metallicity $Z$ is assumed to be equal to the metallicity of the 
stellar population for objects of any gas-to-stars morphology.

For the objects without H$\alpha$ emission within the area of radiation 
from stars (class~0) we assume $A(B)=A(B)_{\rm Gal}+A(B)_{\rm in}$.

The gas contribution $I_B({\rm gas})/I_B({\rm total})$ is assumed to be 0 
for the class~0 objects.

In the catalogue, we present 'true' colours and absolute magnitudes only 
for objects of classes~2 and 0. In doing so, 'true' colours and absolute 
magnitudes for objects with absence of gas emission within the area of 
radiation from stars are equal to colours and magnitudes, corrected for 
the Galactic extinction and inclination effects. We did not calculate the 
colour index $(V-R)_{\rm true}$ for stellar groupings with extremely high 
contribution of gas emission in the $R$ band, 
$I_R({\rm gas})/I_R({\rm total})>0.4$.

Masses and ages of stellar groupings were estimated both for objects 
of classes~2 and 0. We did not include masses and (or) ages for some 
groupings, for which estimates of $m$ and (or) $t$ were obtained with low 
accuracy ($\Delta m/m\ge1$, $\Delta t/t\ge1$ if $t+\Delta t\le10$~Myr).

\subsection{Physical parameters of stellar population in star formation 
regions}
\label{sect:phys}

\begin{figure}
\vspace{5mm}
\resizebox{1.00\hsize}{!}{\includegraphics[angle=000]{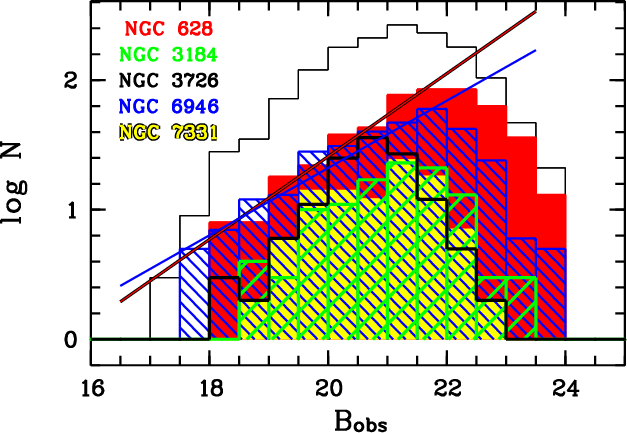}}
\caption{Luminosity functions for star formation regions using apparent 
$B$ magnitude for our full sample (thin black), for objects in NGC~628 
(red), NGC~3184 (green), NGC~3726 (thick black), NGC~6946 (blue), and 
NGC~7331 (yellow). Black-red-black and blue straight lines represent the 
power-law fit of the form of equation~(\ref{equation:lf2}) for samples 
of star formation regions in NGC~628 and NGC~6946, respectively. See the 
text for details.
}
\label{figure:bvis}
\end{figure}

The completeness limits for object samples differ from galaxy to galaxy 
(Fig.~\ref{figure:bvis}), since the observations of galaxies were carried 
out with different telescopes and with different total exposures. In the most 
deeply exposed NGC~628 and NGC~6946, the sample is complete up to apparent 
$B$ magnitudes of $\approx22.0$ and $\approx21.7$~mag, respectively 
(Fig.~\ref{figure:bvis}). The remaining galaxies in the sample were taken 
with lower exposures. The total distribution of 1510 stellar groupings has 
a maximum at $\approx21.3$~mag (Fig.~\ref{figure:bvis}). For galaxies 
with the worst signal-to-noise ratio, object samples are complete up to 
$m(B)\approx20$~mag.

We constructed the luminosity function for stellar groupings in 
the galaxies with the largest numbers of identified star formation 
regions, NGC~628 and NGC~6946. We used a standard power-law luminosity 
function of the form 
\begin{equation}
dN(L_{m(B)})/dL_{m(B)} = \beta L_{m(B)}^{\alpha}
\label{equation:lf}
\end{equation}
which was converted to the form
\begin{equation}
\log N = a\times m(B)+b
\label{equation:lf2}
\end{equation}
for the fitting, where the variables $\alpha$, $\beta$ 
in equation~(\ref{equation:lf}) and $a$, $b$ in 
equation~(\ref{equation:lf2}) are related as $\alpha = -2.5a-1$ and 
$\beta = 2.5(\ln 10)^{-1}10^{b+4.8a}$, respectively.

\begin{figure}
\vspace{5mm}
\resizebox{1.00\hsize}{!}{\includegraphics[angle=000]{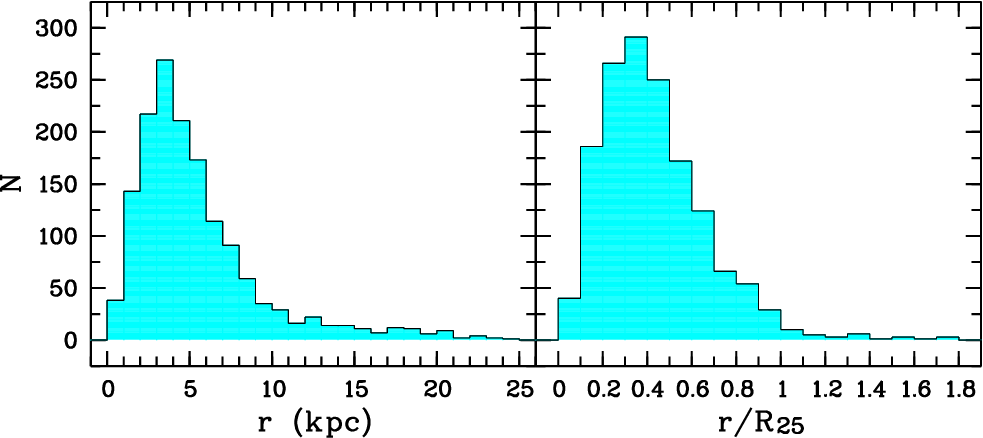}}
\caption{Distributions of studied star formation regions over their 
absolute galactocentric distances (left) and galactocentric distances 
normalized to the disc isophotal radius $R_{25}$ of a galaxy (right) 
for 19 spiral galaxies under study.
}
\label{figure:rad}
\end{figure}

\begin{figure}
\vspace{5mm}
\resizebox{1.00\hsize}{!}{\includegraphics[angle=000]{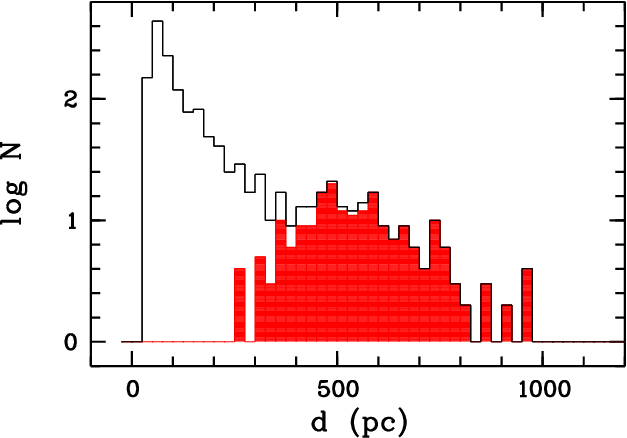}}
\caption{Distribution of studied star formation regions over their 
diameters. The red histogram shows the size distribution of objects in 
galaxies with distances $d>30$~Mpc.
}
\label{figure:size}
\end{figure}

The constructed star formation region luminosity functions are shown in 
Fig.~\ref{figure:bvis}. The luminosity functions have slopes 
$\alpha=-1.80\pm0.05$ for NGC~628 and $\alpha=-1.69\pm0.05$ for NGC~6946. 
These slopes are close to typical ones $\sim-2$ for H\,{\sc ii}~regions 
and young open clusters in spiral galaxies 
\citep{larsen2002,piskunov2006,grijs2006,haas2008,mora2009,chandar2010,
whitmore2010,baumgardt2013,konstantopoulos2013,fouesneau2014,messa2018,
santoro2022}. In particular, \citet{santoro2022} obtained 
$\alpha = -1.7\pm0.1$ for H~\textsc{ii} regions in NGC~628 from the 
integral field-spectroscopy (as part of PHANGS-MUSE survey; 
\citealt{Emsellem2022}), that is in agreement with our measurements based 
on the archival long-slit data.

\begin{figure}
\vspace{5mm}
\resizebox{1.00\hsize}{!}{\includegraphics[angle=000]{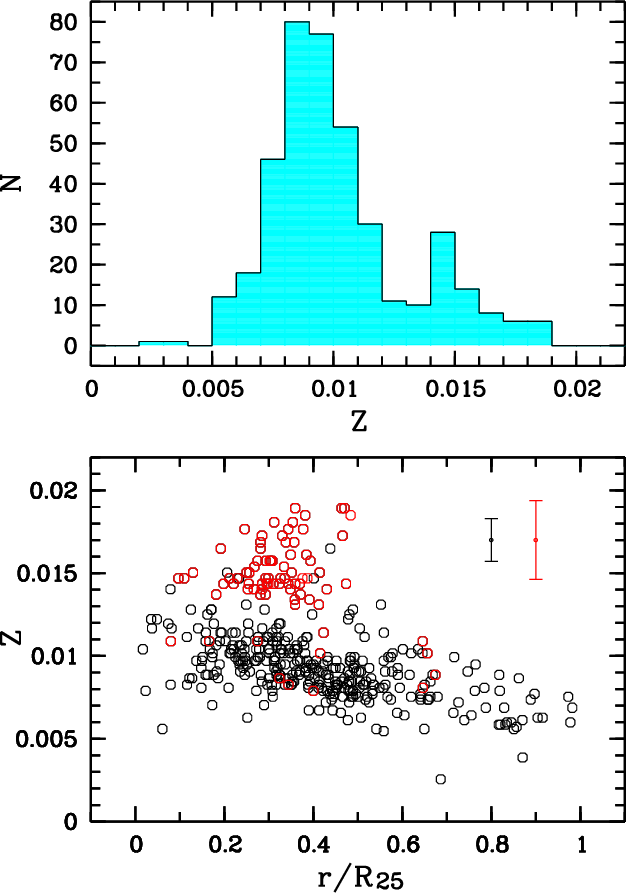}}
\caption{Metallicity distribution of star formation regions (top) and 
radial distribution of metallicities in discs of galaxies (bottom). 
H\,{\sc ii}~regions, in which the abundances were measured with 
H\,{\sc ii}-ChiMistry method, are shown by red circles. Mean error 
bars are given. See the text for details.}
\label{figure:z}
\end{figure}

Most of young stellar groupings are located, as expected, in regions of 
the developed spiral structure at galactocentric distances 
$0.1-0.7 r/R_{25}$ (Fig.~\ref{figure:rad}). At the same time, we 
identified 22 young objects at distances of $1.10-1.74 r/R_{25}$. 
Six of them are located in the irregular galaxy NGC~5585. The remaining 
sixteen are along the minor axis of the highly inclined disc of NGC~7721, 
and the accuracy of finding their galactocentric distances is low. We 
believe that the inclination of the disc of NGC~7721 ($81\degr$) is 
overestimated in the LEDA catalogue. The histogram of the distribution of 
stellar groupings by absolute distances to the centre (left panel of 
Fig.~\ref{figure:rad}) is similar in shape to the distribution in the 
right panel of the figure. This is a consequence of the fact that  
four galaxies with the largest number of identified star forming regions 
(NGC~628, NGC~3184, NGC~3726, and NGC~6946) have close sizes, their 
$R_{25}=10.9-13.3$~kpc (see Table~\ref{table:sample}).

\begin{figure}
\vspace{5mm}
\resizebox{1.00\hsize}{!}{\includegraphics[angle=000]{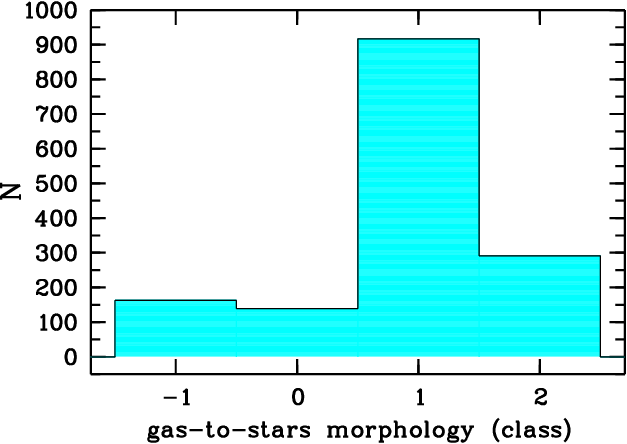}}
\caption{Distribution of studied star formation regions over their 
gas-to-stars morphology. See the description of the catalogue's 
column~(28) in Section~\ref{sect:catalog} for details.
}
\label{figure:hadata}
\end{figure}

The size distribution of stellar groupings is strongly influenced by 
selection effects associated with the fact that the galaxies of our 
sample are located in a wide range of distances from the Milky Way. 
Among nearby galaxies ($d<30$~Mpc), the size distribution of stellar 
groupings has a power-law form with a maximum at $\approx70$~pc 
(Fig.~\ref{figure:size}). Although resolution effects also play a role 
here, this diameter, $70$~pc, is typical for stellar associations 
\citep*{efremov1987,efremov1989,ivanov1991,efremov1995,efremov1998,
elmegreen1994,elmegreen1996,elmegreen2009,odekon2008,marcos2009}, 
and the power law of the size function for stellar associations and 
H\,{\sc ii}~regions is well known 
\citep*{elmegreen2003a,elmegreen2003b,elmegreen2006}. In distant galaxies, 
where we cannot resolve individual associations, the size distribution 
has a maximum of $d=500-600$~pc, which is a typical size of star 
complexes \citep*{efremov1995,efremov1998,zhang2001,elmegreen2000,
elmegreen2002,elmegreen2011}.

Gas in most of the studied H\,{\sc ii}~regions has a sub-solar metallicity, 
$Z\sim0.01 \simeq  0.55 Z_\odot$ (Fig.~\ref{figure:z}). Because we had to 
use different empirical calibration methods (see Section~\ref{sect:spectra}), 
there are systematic discrepancies between the measured metallicity values. 
Whereas R, S, O3N2, and NS-calibrations are in a good agreement with each 
other, H\,{\sc ii}-ChiMistry method gives, on average, 0.1-0.2~dex higher 
values of O/H (see the bottom panel of Fig.~\ref{figure:z}), close 
to solar ones.

We note, however, that errors in determining the chemical abundance 
have small effect on estimates of the age and mass of stellar groupings, 
since the difference in luminosities and colours for evolutionary 
sequences of different metallicities does not exceed the typical 
errors in the measured 'true' luminosities and colour indices of stellar 
groupings (see the sequences in the colour-magnitude and 
colour-colour diagrams below).

The radial distribution of the metallicities of H\,{\sc ii}~regions in 
Fig.~\ref{figure:z} shows a gradient, typical for discs of spiral 
galaxies \citep*{pilyugin2014}.

Among the sample of our objects with available data in the H$\alpha$ 
line (1347 out of 1510), the majority (917 out of 1347, or $68\%$) are 
star formation regions in which photometric (stellar) radiation centre 
is displaced from gas emission centre (class~1), 291 ($22\%$) objects 
are H\,{\sc ii}~regions in which optical radiation from stars coincides 
with ionized gas emission (class~2), and 139 ($10\%$) objects have no 
gas emission within the area of optical radiation from stars 
(class~0; Fig.~\ref{figure:hadata}). Evolutionary classification scheme 
of \citet{whitmore2011} predicts objects of class~1 to be between 
4-5 and 8-10~Myr old, and objects of class~2 to be younger than 4-5~Myr. 
Taking into account that the youngest star formation regions with an age 
of 1-2~Myr are not visible in optics due to high extinction in the 
surrounding gas-dust cloud \citep[e.g.][]{Kim2021,Kim2023}, as well as 
selection effects, due to which younger and dusty objects have a larger 
$m(B)$ than objects of class~1 of the same luminosity, a ratio of 3:1 for 
stellar groupings of class~1 and class~2 seems reasonable.

\begin{figure}
\vspace{5mm}
\resizebox{1.00\hsize}{!}{\includegraphics[angle=000]{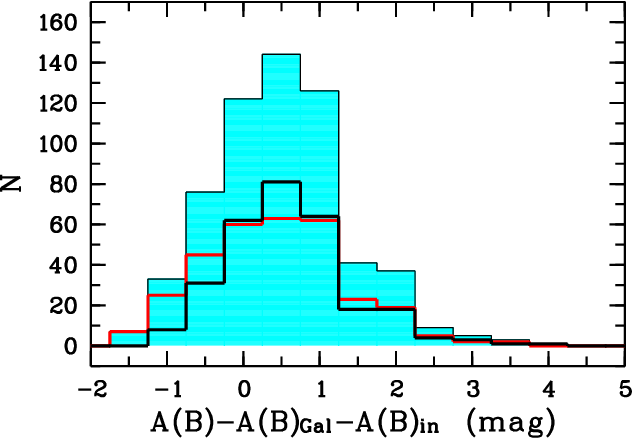}}
\caption{Frequency distribution of star formation regions over intrinsic 
extinction within the star formation regions for objects of all classes 
(cyan), for objects of class~2 (black), and for objects of class~1 (red). 
Intrinsic extinction within the star formation region is determined as 
a difference between the extintion found from Balmer decrement, $A(B)$, and 
sum the of the Galactic extinction and the dust extinction due 
to the inclination of a galaxy, $A(B)_{\rm Gal}+A(B)_{\rm in}$. See the 
text for details.
}
\label{figure:ab}
\end{figure}

Spectral data are not available for the every object from our sample; 
therefore, we were able to obtain the Balmer decrement and estimate the 
extinction $A(B)$ in star formation regions only for 604 objects in the 
catalogue. We present in Fig.~\ref{figure:ab} the distribution of star 
formation regions by intrinsic extinction computed from Balmer 
decrement, $A(B)$, and corrected for the Galactic extinction 
and the dust extinction due to the inclination of a galaxy, 
$A(B)_{\rm Gal}+A(B)_{\rm in}$. Typical extinction 
$A(B)-A(B)_{\rm Gal}-A(B)_{\rm in}\approx0.5$~mag in H\,{\sc ii}~regions. 
For some regions, it can reach 4~mag, but usually does 
not exceed 2~mag (Fig.~\ref{figure:ab}). Note that among the regions, 
for which the Balmer decrement $A(B)<A(B)_{\rm Gal}+A(B)_{\rm in}$, the 
regions with a displaced gas emission centre (class~1) dominate. 
Among the regions of class~2, we found only 47 objects ($16\%$), in which 
the negative $A(B)-A(B)_{\rm Gal}-A(B)_{\rm in}$ exceeds the errors 
$\Delta A(B)$. Note that the mean $\Delta A(B)=0.26$~mag for class~2 
objects and $0.32$~mag for class~1 objects.

Negative $A(B)-A(B)_{\rm Gal}-A(B)_{\rm in}$, as a rule, are not 
the result of errors in spectroscopic measurements. $A(B)_{\rm in}$ is 
the average value for the galaxy as a whole. However, extinction in a 
galaxy is not a constant value, it has radial and vertical gradients, 
as well as local variations. Therefore, $A(B)_{\rm in}$ for a particular 
cluster depends on its galactocentric distance and vertical distance 
from the disc plane. It may be less than the average for the galaxy as 
a whole.

We note specific cases when the $A(B)_{\rm Gal}$ extinction can also 
differ from the average for an extragalactic cluster. An example is 
the galaxy NGC~6946 (see Table~\ref{table:sample}), located at a low 
Galactic latitude ($b=11.7\degr$). Small local changes in extinction 
in the Milky Way can give significant deviations of $A(B)_{\rm Gal}$ 
over the field of the galaxy \citep*{efremov2011}.

\begin{figure}
\vspace{5mm}
\resizebox{1.00\hsize}{!}{\includegraphics[angle=000]{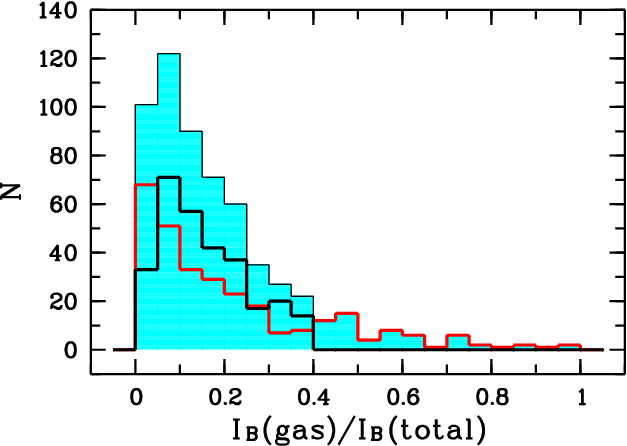}}
\caption{Distributions of studied star formation regions over nebular 
emission contribution in the $B$ band. The symbols are the same as in 
Fig.~\ref{figure:ab}.
}
\label{figure:gas}
\end{figure}

The distribution of objects by gas emission contribution to the total 
radiation in photometric broadbands is similar to what we obtained in 
\citet{gusev2016} in fig.~11. The characteristic nebular contribution in 
the $B$ band is about $10\%$ (Fig.~\ref{figure:gas}). Among regions of 
class~1, a relative excess of objects with small ($<5\%$) and 
large ($>40\%$) nebular contributions is observed. Apparently, in the 
first case, we have spectral observations, in which the slit passed 
through the photometric (stellar) centre of H\,{\sc ii}~region, and in 
the second case, it passed through the centre of gas emission.

\begin{figure}
\vspace{5mm}
\resizebox{1.00\hsize}{!}{\includegraphics[angle=000]{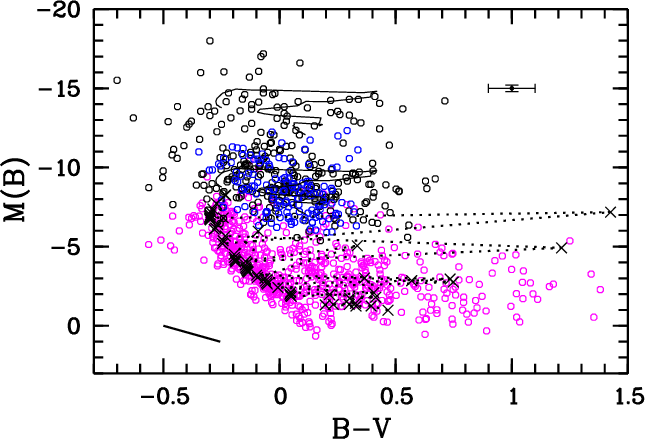}}
\caption{True colours and luminosities of studied young stellar groupings 
(black and blue open circles) and open clusters in the Milky Way 
(magenta open circles) compared with the {\it Standard} mode of SSP models 
(continuously populated IMF; black curves) and the {\it Extended} 
SSP mode (randomly populated IMF; black crosses connected by dotted 
lines). The objects of class~2 are indicated by black, and the objects of 
class~0 are shown by blue. Mean error bar is given. Two evolutionary 
sequences of the {\it Standard} mode with an adopted characteristic 
metallicity of $Z=0.008$, drawn in the age interval from 1 to 100~Myr 
are shown. The sequences were computed for the masses of star clusters of  
$1\cdot 10^6 M_{\odot}$ and $1\cdot 10^4 M_{\odot}$. The sample of 
{\it Extended} SSP models with randomly populated IMF generated for the 
characteristic metallicity of galactic open clusters ($Z = 0.019$), shown 
with crosses connected by dotted lines in order of increasing age, 
demonstrates the evolutionary sequence for $500 M_{\odot}$ in the 
age range of 1~Myr to 1~Gyr.
}
\label{figure:cmd}
\end{figure}

We present a colour-magnitude diagram $B-V$ versus $M(B)$ for the 'true' 
colours and luminosities of stellar population of the studied star 
formation regions of classes~0 and 2, as well as open star clusters in 
the Milky Way from the catalogue of 
\citet{kharchenko2005a,kharchenko2005b,kharchenko2009} in 
Fig.~\ref{figure:cmd}. As can be seen from the figure, the vast majority 
of objects are well described by synthetic evolutionary SSP 
sequences for continuously and randomly populated IMF.

Note that among young stellar groupings without H$\alpha$ emission 
(class~0) there are no high-mass objects with 
$M>2\cdot10^5 M_{\odot}$. The reason for this phenomenon will be 
discussed in Section~\ref{sect:discus}.

Figure~\ref{figure:cmd} shows that the brightest open star clusters in our 
Galaxy and the dimmest stellar groupings from our sample are superimposed on 
each other in the colour-magnitude diagram. This confirms the conclusion 
of \citet{gusev2016} that extragalactic young stellar groupings and open 
star clusters in the Milky Way form a continuous sequence of masses and 
ages and they represent a single evolutionary sequence of objects at 
different stages of their evolution.

\begin{figure}
\vspace{10mm}
\resizebox{0.99\hsize}{!}{\includegraphics[angle=000]{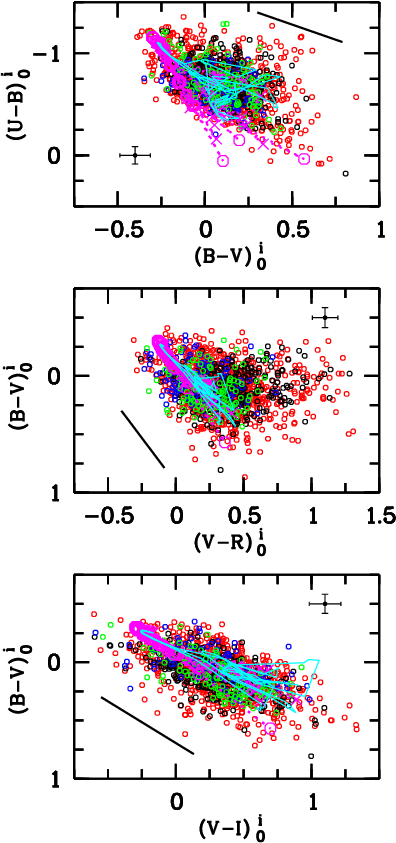}}
\caption{Colour-colour diagrams $(U-B)_0^i-(B-V)_0^i$, 
$(B-V)_0^i-(V-R)_0^i$, and $(B-V)_0^i-(V-I)_0^i$ for star formation 
regions of different gas-to-stars morphology. The class~2 objects 
are shown by black, the class~1 objects are given by red, the class~0 
objects are indicated by blue, and the objects with no H$\alpha$ data 
are shown by green. The colour indices are corrected for the Galactic 
extinction and the dust extinction due to the inclination of a 
galaxy. Mean error bars are given. The black straight line in the 
corner of the diagrams is parallel to the extinction vector. Cyan curves 
show SSP models with continuously populated IMF with $Z=0.008$, 0.012, and 
0.018 for the age interval from 1 to 100~Myr. Circles and crosses 
connected by magenta dotted lines, in order of increasing age, indicate 
two samples of {\it Extended} SSP models with randomly populated IMF with 
$Z=0.008$ generated for star cluster masses $5\cdot 10^3 M_{\odot}$ 
(crosses) and $500 M_{\odot}$ (circles) in the age range of 1~Myr to 1~Gyr.
}
\label{figure:color}
\end{figure}

We present colour-colour diagrams of colour indices, corrected for the 
Galactic extinction and the dust extinction due to the 
inclination of a galaxy, for all objects of our sample in 
Fig.~\ref{figure:color}. As can be seen from the figure, most of the star 
formation regions are well superimposed on the evolutionary sequences 
of young stellar systems with $t\le10$~Myr. The exception is the 
$(B-V)_0^i-(V-R)_0^i$ diagram, where a 'tail' of objects with an 
anomalously large $(V-R)_0^i$ is observed. The excess radiation in the 
$R$ band is due to the large contribution of gas emission lines to this 
photometric band.

The star formation regions with no gas emission (class~0) and the 
regions, for which there are no H$\alpha$ data, lie more compactly along 
the evolutionary sequences on the colour-colour diagrams, than the 
regions with the presence of H$\alpha$ emission (classes~1 and 2). This is 
due to selection effects: objects without gas emission were selected based 
only on their colour indices $(U-B)_0^i$ and $(B-V)_0^i$ (see 
Section~\ref{sect:select}). The largest scatter in the colour-colour 
diagrams is observed for regions with displaced gas emission centre 
(class~1), because the gas emission contribution and 'true' (Balmer) 
absorption in the H\,{\sc ii}~region are not taken into account here.

\begin{figure}
\vspace{10mm}
\resizebox{0.99\hsize}{!}{\includegraphics[angle=000]{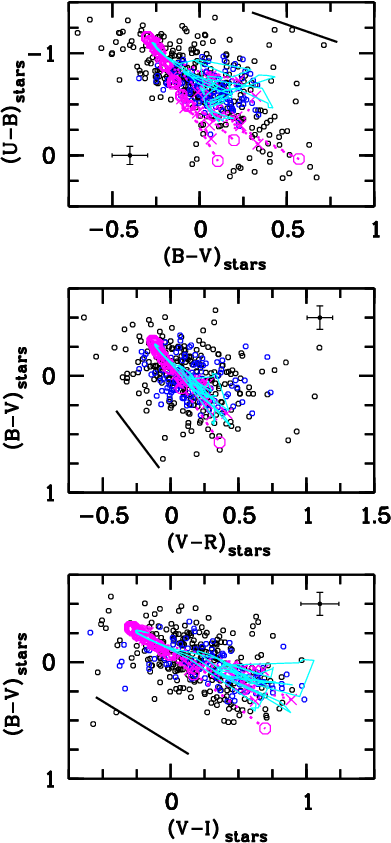}}
\caption{Colour-colour diagrams $(U-B)-(B-V)$, $(B-V)-(V-R)$, and 
$(B-V)-(V-I)$ for true colours of studied young stellar groupings. 
The class~2 objects are shown by black open circles and the class~0 objects 
are indicated by blue open circles. Other symbols are the same as 
in Fig.~\ref{figure:color}.
}
\label{figure:color2}
\end{figure}

On colour-colour diagrams showing the 'true' colours of stellar groupings 
(Fig.~\ref{figure:color2}), objects are located more compactly along 
evolutionary sequences than the star formation regions in the diagrams 
$(U-B)_0^i-(B-V)_0^i$, $(B-V)_0^i-(V-R)_0^i$, and $(B-V)_0^i-(V-I)_0^i$ 
(Fig.~\ref{figure:color}). This may indicate the correctness of our 
estimates of the nebular emission contribution and extinction 
calculated from the Balmer decrement.

Part of stellar groupings of class~2 with $(B-V)_{\rm stars}\sim 0.4$ 
and $(U-B)_{\rm stars}>0$ is well described only by evolutionary 
sequences with a randomly populated IMF in the $(U-B)-(B-V)$ diagram. 
In the $(U-B)_0^i-(B-V)_0^i$ diagram, there are no star formation regions 
with such $(U-B)_0^i$ and $(B-V)_0^i$ (Fig.~\ref{figure:color}).

Stellar groupings with gas emission (class~2) have systematically smaller 
colour indices $U-B$ and $B-V$ than regions without gas emission (class~0). 
This is especially clearly seen in the $(U-B)-(B-V)$ diagram
(Fig.~\ref{figure:color2}). This is an expected result, reflecting the 
fact that star formation regions with gas emission should be younger and 
bluer on average.

\begin{figure}
\vspace{5mm}
\resizebox{1.00\hsize}{!}{\includegraphics[angle=000]{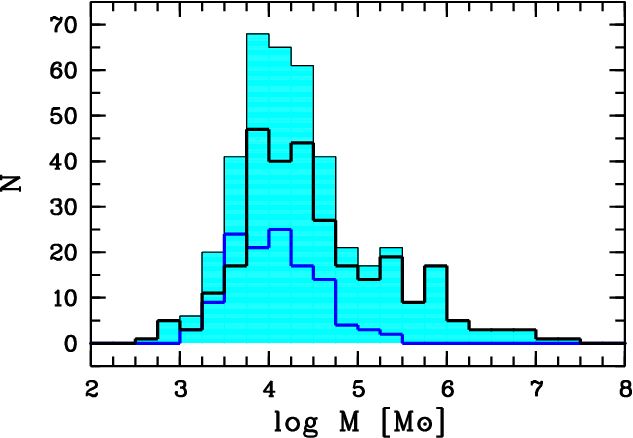}}
\caption{Frequency distribution of stellar groupings over mass for all 
regions (cyan), for class~2 objects (black), and for class~0  objects (blue).
}
\label{figure:mass}
\end{figure}

Amongst 430 stellar groupings of classes~0 and 2, we were able to 
estimate the mass for 409 and the age for 391 objects using evolutionary 
models. Most star clusters have masses in the range of 
$3\cdot10^3-3\cdot10^5 M_{\odot}$ (Fig.~\ref{figure:mass}). Two thirds of 
stellar groupings can be attributed to massive star clusters with 
$M>1\cdot10^4 M_{\odot}$. The minimum masses were fixed for stellar 
groupings no.~465 in NGC~628 ($430 M_{\odot}$) and nos.~286, 301, 367 
in NGC~628, and no.~634 in NGC~3184 ($790 M_{\odot}$). According to our 
estimates, stellar complexes no.~1439 in NGC~7678 and no.~890 in 
NGC~5351 have the maximum masses ($3\cdot10^7$ and 
$1.3\cdot10^7 M_{\odot}$, respectively).

As noted above, among the objects without gas emission (class~0) 
there are no high-massive star complexes with $M>2.2\cdot10^5 M_{\odot}$. 
We also did not find stellar groupings of class~0 with masses 
$M<1.1\cdot10^3 M_{\odot}$ (Fig.~\ref{figure:mass}).

\begin{figure}
\vspace{5mm}
\resizebox{1.00\hsize}{!}{\includegraphics[angle=000]{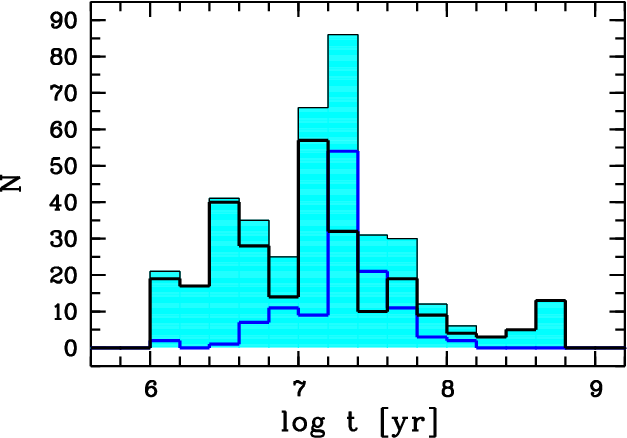}}
\caption{Frequency distribution of stellar groupings over age. The 
symbols are the same as in Fig.~\ref{figure:mass}.
}
\label{figure:t}
\end{figure}

The age range of stellar groupings turned out to be unexpectedly 
wide: from 1 to 560~Myr (Fig.~\ref{figure:t}). 154 regions ($39\%$) are 
younger than 10 Myr, another 137 objects ($35\%$) have an age of 
$10-25$~Myr.

Objects without gas emission (class~0), as expected, turned out to be on 
average older than regions with H$\alpha$ emission. The boundary, at 
which regions of class~0 begin to predominate, is the age of $15-16$~Myr 
(Fig.~\ref{figure:t}). Objects without H$\alpha$ emission are practically 
absent among very young ($t<4$~Myr) and relatively old ($t>130$~Myr) 
stellar groupings (Fig.~\ref{figure:t}). A probable reason for the 
presence of gas emission in stellar groupings older than 10~Myr is 
prolonged or multi-burst star formation, which is poorly described in 
terms of SSP evolutionary models (see Section~\ref{sect:discus} for 
details).

\begin{figure*}
\centering
\includegraphics[width=\linewidth]{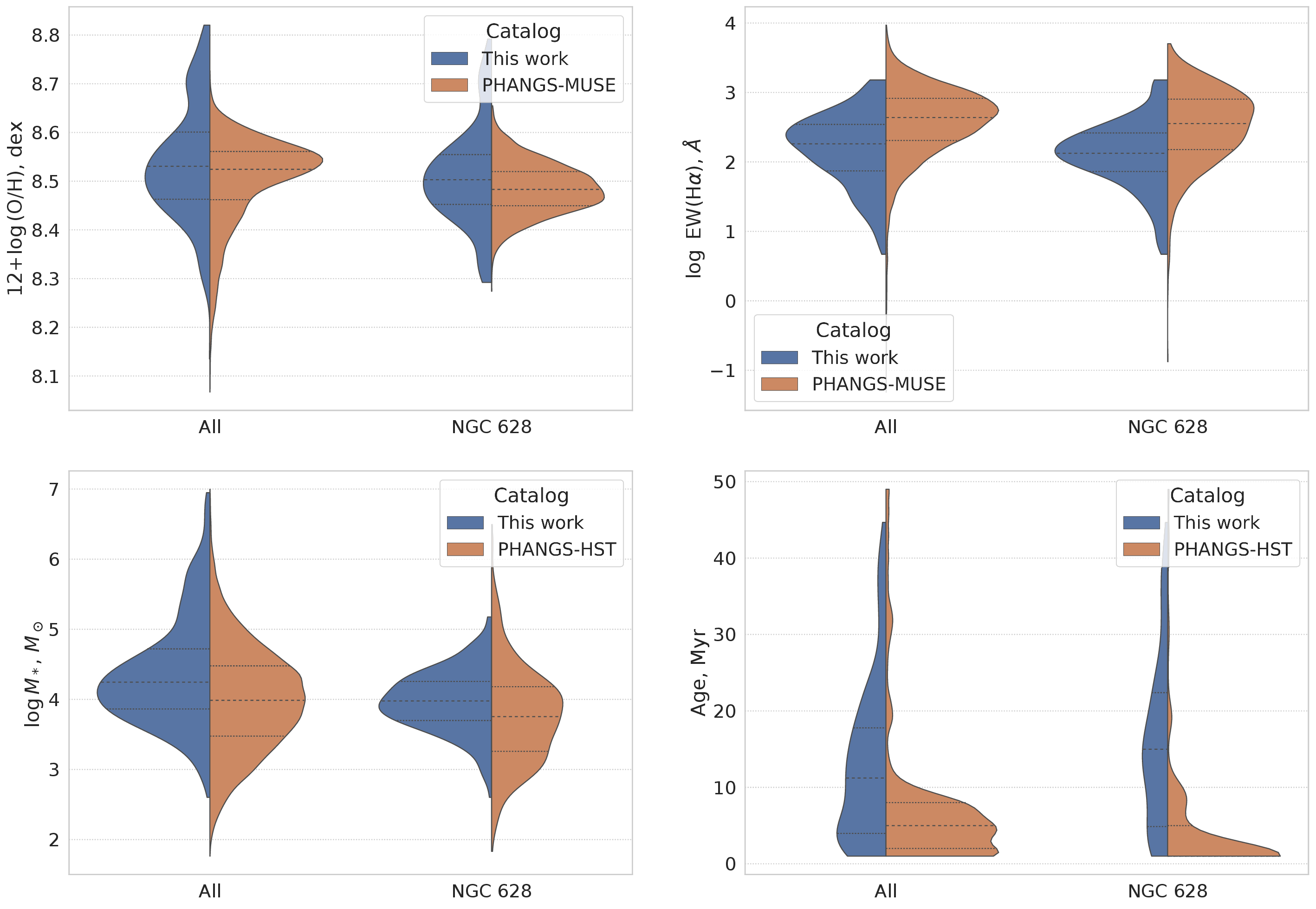}
\caption{Distribution of the gas-phase oxygen abundance (top-left 
panel), equivalent width of H$\alpha$ line (top-right panel), stellar 
mass (bottom-left panel) and age (bottom-right panel) derived for stellar 
groupings from this study (blue colour) in comparison to the distribution 
of the same parameters for H~\textsc{ii} regions and compact star clusters 
and young associations for the sample of other nearby galaxies from the 
PHANGS survey based on MUSE and HST observations (taken from the 
catalogues by \citealt{Whitmore2021, Thilker2022, Groves2023, Larson2023}. 
The right-hand set of histograms on each plot shows the distribution for 
NGC~628 -- the only overlapping object in our samples. The values of 
EW(H$\alpha$) from PHANGS-MUSE were corrected for the background old 
stellar population, while no such correction was performed for our 
sample (see text).
}
\label{fig:phangs}
\end{figure*}

In order to verify that the measured properties of the ionized gas 
and stars presented in our catalogue are consistent with what is 
typically observed in nearby galaxies by other authors, we compare 
their distribution to what is derived from PHANGS-MUSE \citep{Emsellem2022} 
and PHANGS-HST \citep{Lee2022} data. Within PHANGS survey, 19 nearby 
galaxies were mapped with MUSE and 38 galaxies -- with HST, while only 
one galaxy (NGC~628) is also in our catalogue. From these data, the 
properties of about 30000 H~\textsc{ii} regions \citep{Groves2023} and 
about 100000 young compact star clusters \citep{Whitmore2021,Thilker2022} 
and OB associations \citep{Larson2023} were derived, and currently the 
corresponding catalogues are ones of the most comprehensive sources of 
the resolved observational properties of the H~\textsc{ii} regions and 
star clusters in nearby galaxies. In Fig.~\ref{fig:phangs} we show the 
distribution of the gas-phase oxygen abundances, EW(H$\alpha$), total 
stellar mass and age of the star groupings in our study (blue histograms), 
and those taken from the PHANGS catalogues (orange histogram) derived from 
the MUSE (for oxygen abundance and equivalent width) and HST (for stellar 
mass and age) observations. As follows from these plots, the regions from 
our catalogue cover roughly the same range of metallicities and stellar 
masses, although the PHANGS-HST data are more sensitive and complete at 
the low-mass range of star clusters. The fraction of the relatively old 
star clusters is significantly higher in our sample, probably due to the 
fact that we are studying the larger stellar groupings (thus their average 
age can be older than the age of the youngest individual compact clusters 
there), while PHANGS-HST resolves more compact star clusters and young 
stellar associations. Finally, the values of EW(H$\alpha$) measured for 
our sample are slightly lower than for PHANGS-MUSE H~\textsc{ii} regions. 
We note here that the latter values were corrected for contamination by 
background old stellar population, which is quite strong in the IFU data 
like PHANGS-MUSE. \citet{Scheuermann2023} showed that background-corrected 
values of EW(H$\alpha$) from PHANGS-MUSE catalogue are about an order of 
magnitude larger than the observed ones, and in Fig.~\ref{fig:phangs} we 
show the corrected values from that paper. Our measurements rely mostly 
on the long-slit data and less suffer from this effect, although slight 
displacement between the distributions can be due to the fact that we 
did not perform such correction for our measurements (but also due to 
in general older ages of the stellar population).

\section{Discussion}
\label{sect:discus}

The duration of star formation is approximately proportional to the mass 
of a molecular cloud and a star grouping formed from it. Star formation 
in massive star complexes lasts $\sim20$~Myr \citep{efremov1998}. 
Probably, in the most massive star complexes, we observe the emission of 
hydrogen from the last, recent starburst. At the same time, the first 
starburst could have occurred relatively long ago. In the absence of 
a recent burst of star formation, the colour indices of such 
complexes are not well defined by SSP models as several generations of 
stars could be observed at the same place. Therefore, we have not 
identified any star complexes without gas emission (class~0) with a mass 
$M>2.2\cdot10^5 M_{\odot}$ (see Figs.~\ref{figure:cmd}, 
\ref{figure:mass}).

The relative duration of star formation in large star complexes can 
also explain the fact that there are no low-mass stellar 
systems ($M<10^4 M_{\odot}$) among the 'oldest' stellar groupings 
($t>130$~Myr) of any classes (2 and 0).

\begin{figure}
\vspace{5mm}
\resizebox{1.00\hsize}{!}{\includegraphics[angle=000]{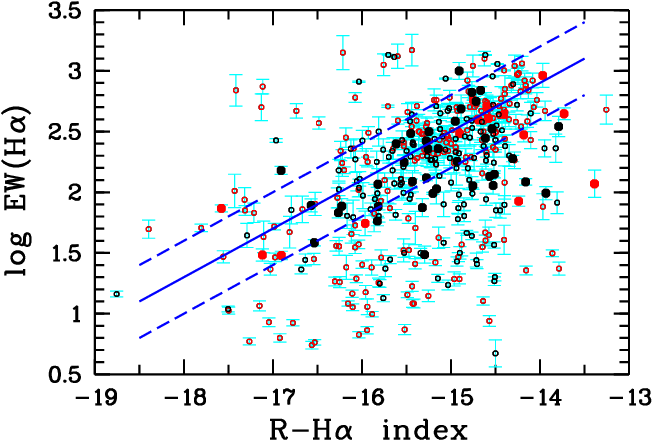}}
\caption{$R-$H$\alpha$ index versus $\log$~EW(H$\alpha$) diagram for 
star formation regions of class~2 (open black circles) and~1 (open red 
circles) with $\Delta\log$~EW(H$\alpha)<0.15$~dex. Our objects from 
\citet{gusev2016} are shown as large filled circles. Error bars are 
given. The solid line is a linear fit from equation~(\ref{equation:rha}). 
Dashed lines are upper and lower $1\sigma$ errors of 
$\log$~EW(H$\alpha$) from equation~(\ref{equation:rha}).
}
\label{figure:rha}
\end{figure}

As we noted in Introduction, EW(H$\alpha$) appears to be one of the 
commonly used age indicators. The definite advantage of this indicator 
is its insensitivity to the interstellar extinction. 
In Fig.~\ref{figure:rha} we compare EW(H$\alpha$) to the 
$R-{\rm H}\alpha$ index (also independent on reddening) that we 
introduced in Section~\ref{sect:catalog}. From the definition of the 
index, $\log {\rm EW(H}\alpha)\sim0.4(R-{\rm H}\alpha)$, and the 
meaning of that index is the same as of EW(H$\alpha$) assuming the 
underlying stellar continuum has a flat shape. Formally found linear 
regression coefficient between these two indexes for the considered 
star formation groupings is equal to 0.407, very close to 0.4. 
The resulting relation between the indexes can be described as
\begin{equation}
\log {\rm EW}({\rm H}\alpha) = 0.4(R-{\rm H}\alpha)+(8.5\pm0.3).
\label{equation:rha}
\end{equation}
However, the scatter in Fig.~\ref{figure:rha} and the derived 
uncertainties of the offset in equation~\ref{equation:rha} 
($\pm0.3$~dex) do not allow its accurate application to the real 
data. In particular, there is a numerous group of H\,{\sc ii}~regions, 
mostly of class~1, with anomalously low EW(H$\alpha$) for a given 
$R-{\rm H}\alpha$ index (Fig.~\ref{figure:rha}). 
Note that the mean error $\Delta \log {\rm EW(H}\alpha)$ 
is equal to 0.06~dex for the objects in Fig.~\ref{figure:rha}.

Analyzing the EW(H$\alpha$), obtained by different authors for the same 
H\,{\sc ii}~regions, we found that the differences in the measured 
EW(H$\alpha$) can reach a factor of 2-3. This is also why, after 
averaging (see Section~\ref{sect:spectra}), the EW(H$\alpha$) errors 
in Fig.~\ref{figure:rha} and in the catalogue turned out to be larger 
than the EW(H$\alpha$) errors of individual measurements.

In our opinion, there are two main reasons for the large scatter in 
EW(H$\alpha$) measurements. The first one is related to the fact that 
the EW(H$\alpha$) values are sensitive to the choice of the background 
area (continuum from the underlying stellar disc). 
Small changes in the continuum under the H$\alpha$ line lead to 
significant changes in the EW(H$\alpha$) value. The second reason, which 
is specific for slit spectroscopy, is related to the fact that different 
parts of H\,{\sc ii}~region with different EW(H$\alpha$) can fall into 
the slit. Additional reasons for the scatter in EW(H$\alpha$) 
measurements can be also due to variations of the filling factors in 
H\,{\sc ii}~regions and the fraction of ionizing photons which escape 
from the nebulae. Therefore, the use of EW(H$\alpha$) as an age 
indicator in H\,{\sc ii}~regions should be treated with caution.

\begin{figure}
\vspace{5mm}
\resizebox{1.00\hsize}{!}{\includegraphics[angle=000]{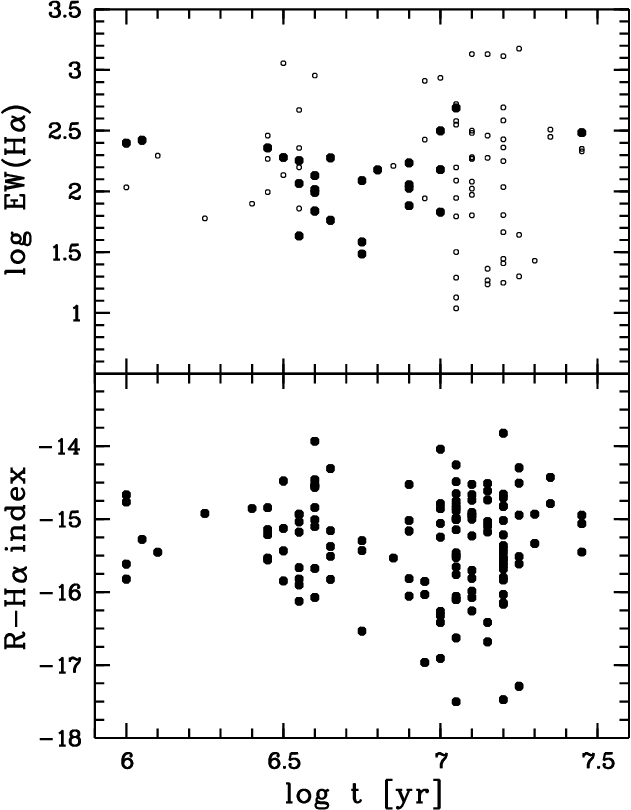}}
\caption{Age versus $\log$~EW(H$\alpha$) (top) and age versus 
$R-$H$\alpha$ index (bottom) diagrams for the most accurately measured star 
formation regions ($\Delta\log$~EW(H$\alpha)<0.15$~dex for the objects 
in the top panel and $\Delta\log t<0.2$~dex). Symbols in the top panel 
are the same as in Fig.~\ref{figure:rha}. See the text for details.
}
\label{figure:trhaew}
\end{figure}

We plotted the age versus EW(H$\alpha$) and the age versus $R-{\rm H}\alpha$ 
index diagrams in Fig.~\ref{figure:trhaew} for the objects with the 
most precisely measured EW(H$\alpha$) and $t$. Starburst99 evolutionary 
models \citep{leitherer1999} show that EW(H$\alpha$) in 
H\,{\sc ii}~regions decreases from $>1000$\AA\ in the youngest star 
clusters to $\approx30-40$\AA\ in the regions with an age of 10~Myr 
\citep{reines2010}. The lowest EW(H$\alpha$) values for a given age 
in Fig.~\ref{figure:trhaew} correspond to those predicted by Starburst99 
models, however, the largest values ($\sim1000$\AA) are found for young 
stellar regions of all ages, up to $t=17$~Myr.

A similar picture is observed in the age versus $R-{\rm H}\alpha$ index 
diagram: excluding the youngest regions with $t\approx1$~Myr, the minimum 
values of the index fall from $-15$ to $-18$ in the age range of 
$3-15$~Myr, while the maximum indices remain constant, 
$R-{\rm H}\alpha=-14$ (Fig.~\ref{figure:trhaew}).

Apparently, the lack of a strong correlation between age and EW(H$\alpha$) 
predicted by evolutionary models is due to the complex star formation 
history in the H\,{\sc ii}~regions, which is poorly described by 
SSP models. This is also indicated by the fact that there are no objects 
with a low $R-{\rm H}\alpha$ index ($<-16.5$) among young low-massive 
($M<10^4 M_{\odot}$) stellar groupings.

At the same time, we believe that the introduced $R-{\rm H}\alpha$ index 
can be used when EW(H$\alpha$) is not available because of absence 
of spectral observations of star formation regions for their analysis. 
Given the weakness of the correlation with the age of the regions 
derived from SED fitting, one cannot rely on either of these indexes for 
the precise age-dating based without properly addressing tow most probable 
sources of the scatter mentioned above (contamination by the background 
stellar population, uncertainties in the escape fraction of the ionizing 
quanta and the differences in the area covered by the aperture). Integral 
field spectroscopy is necessary to overcome some of such limitations 
related to the incomplete coverage of H\,{\sc ii} region. However, even 
then the measured values EW(H$\alpha$) often disagree with those predicted 
from models \citep[e.g.][]{Morisset2016,Kreckel2022} and only weakly 
correlate with the stellar association ages measured from SED fitting 
\citep[e.g.][]{Scheuermann2023}.

Our sample contains various types of young stellar objects: 
OB associations, open clusters, stellar aggregates, and star complexes. 
Their dynamic evolution is different. Large stellar aggregates with sizes 
$>150$~pc and star complexes with $d\approx500-600$~pc have a complex 
structure and contain conglomerates of star clusters and associations. 
H\,{\sc ii}~regions and associations may expand with age inside the 
complex, but the size of the complex depends fundamentally on the physical 
parameters of the surrounding interstellar matter and the magnetic field 
\citep{elmegreen2003a,elmegreen2003b,gusev2013b}.

Modern high-resolution studies using {\it Hubble Space Telescope} data 
of resolved H\,{\sc ii}~regions show that the size of an 
H\,{\sc ii}~region is a function of the age of the stellar population 
\citep{whitmore2011,kim2012}. However, this dependence is observed up to 
an age of $5-6$~Myr and a diameter of 40~pc. The sizes of  
H\,{\sc ii}~regions (future star clusters) starting from 40~pc weakly 
depend on the age of the stellar population \citep{whitmore2011}. However, 
the 'age--size' relation is observed for star associations over a wider 
range of ages and sizes \citep{efremov1998,zwart2010}. Unfortunately, 
as we noted in Introduction, the youngest ($t\le10$~Myr) clusters and 
associations are poorly differentiated by their parameters 
\citep{gieles2011}, and the minimum linear resolution of our observations 
is $30-40$~pc in the nearest galaxies.

\begin{figure}
\vspace{5mm}
\resizebox{1.00\hsize}{!}{\includegraphics[angle=000]{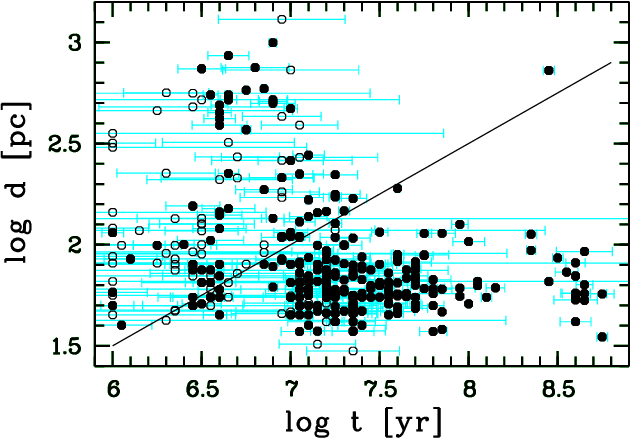}}
\caption{The age versus size diagram for stellar groupings. Objects with 
$\Delta\log t<0.2$~dex are indicated by filled circles. The solid line 
represents the dependence $t\sim d^2$. Error bars are shown. See the text for 
details.
}
\label{figure:td}
\end{figure}

Figure~\ref{figure:td} illustrates the dependence between the age and the 
size of the studied stellar groupings. To make a homogeneous sample, multiple 
(double, triple, and complex) objects have been excluded from the graph 
(see comments for column~(55) of the catalogue). Visually, we do not 
find any correlation between the age and the size of the young stellar 
groupings in the figure. However, we can separate large star complexes, 
a few stellar aggregates, and numerous star clusters. They vary in size 
but do not show 'age--size' dependence. However, the distribution of 
young ($t<10$~Myr) small ($d<120$~pc) stellar groupings on the graph 
seems to indicate the presence of the 'age--size' relation. Larger star 
associations are older. A diffusion-driven expansion, which produces 
a relation $t\sim d^2$ between age and size, seems to play the main 
role here (see Fig.~\ref{figure:td}).

\begin{figure}
\vspace{5mm}
\resizebox{1.00\hsize}{!}{\includegraphics[angle=000]{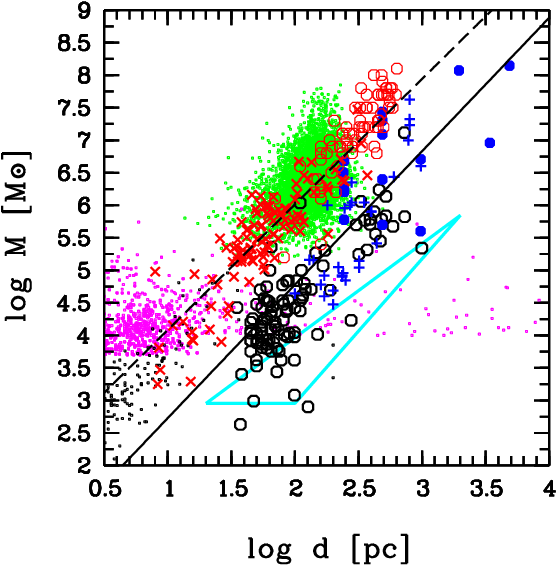}}
\caption{Size versus mass diagram for stellar groupings from our sample 
with $\Delta M/M<0.2$ (black open circles), young massive cluster 
complexes in the very distant ($z\sim1.5$) galaxy Sp~1149 (blue filled 
circles) and cluster complexes in the local galaxies (blue crosses) by 
\citet{adamo2013}, young star clusters in NGC~628, NGC~1313, and 
NGC~5236 \citep{ryon2015,ryon2017} (magenta dots), Small Magellanic 
Cloud star clusters \citep{gatto2021} (black dots), stellar 
conglomerations in NGC~1566 \citep{gouliermis2017} (area inside the 
cian triangle), giant molecular clouds from \citet{bolatto2008} 
(red crosses), \citet*{wei2012} (red open circles), and 
\citet{rosolowsky2021} (green dots). The solid line is a linear fit, 
computed for local and high-$z$ young massive complexes in 
\citet{adamo2013}. The dashed line is a linear fit, computed for 
giant molecular clouds from the \citet{bolatto2008} sample. Mass 
measurement error bars for our objects are smaller than circle sizes 
in the figure. See the text for details.
}
\label{figure:d_m}
\end{figure}

A correlation between sizes and masses of giant molecular 
clouds (GMCs), $M\sim d^2$, was found for the first time by 
\citet{larson1981}. That correlation has been repeatedly confirmed later 
\citep[see, e.g.,][]{hopkins2012}. The mass--size relation for young 
star complexes was found to be close to that of GMCs,
\citep[see][and references therein]{adamo2013}. It reflects the fact 
that young star complexes are the direct descendants of GMCs. Using 
the GMC sample of \citet{bolatto2008} and their own sample of young massive 
clusters, \citet{adamo2013} gave a relation $M\sim d^{2.0\pm0.3}$ for 
young star complexes and $M\sim d^{1.9\pm0.1}$ for GMCs (see 
Fig.~\ref{figure:d_m}). However, more recent studies of star clusters and 
GMCs have shown more complex relations between their masses and sizes 
\citep[see, e.g., compilation in][]{grudic2021}.

The dependence between sizes and masses of stellar groupings 
from our sample is presented in Fig.~\ref{figure:d_m}.  As in the 'age--size' 
diagram (Fig.~\ref{figure:td}), we excluded multiple objects from the 
graph. We also excluded objects with mass estimation errors $>20\%$. 
The stellar groupings from our sample are in fairly good agreement with the 
dependence $M\sim d^2$ obtained by \citet{adamo2013} for the size range 
from 50 to 1000~pc (Fig.~\ref{figure:d_m}). Note that small stellar 
groupings ($d=50-100$~pc) also fit well with the 'size--mass' dependence 
(\citet{adamo2013} studied star clusters larger than 100~pc). At 
the same time, our data also agree well with the results of 
\citet{gouliermis2017} for stellar conglomerations with the highest 
surface brightness from their sample (upper side of the cyan triangle in 
the figure).

The 'size--mass' relation which is observed for star clusters, associations 
and complexes is a relic of the same dependence for their ancestors, the 
GMCs. The vertical shift between the GMCs and the stellar groupings in 
the diagram is due to the star formation efficiency: only a fraction 
of the gas in the GMCs will form stars \citep{bastian2005,adamo2013}. 
The width of the band occupied by stellar groupings along the dependence 
line $M\sim d^2$ testifies to the differing efficiency of star formation 
in different regions, which is usually $1-7\%$.

\section{Conclusions}

In this paper, we present the results of analysis of the catalogue comprising 
parameters of 1510 young stellar groupings.This catalogue is based on the 
combination of spectroscopic, photometric and H$\alpha$ spectrophotometric 
data for star formation regions in 19 galaxies. We have studied the 
morphology of stellar groupings and their relation to the associated 
H$\alpha$ emission region. Extinctions for 743, metallicities for 402, 
ages for 391, and masses for 409 stellar groupings were estimated.

We used a continuously populated IMF for high-massive clusters 
($M>1\cdot10^4 M_{\odot}$) and a randomly populated IMF for star clusters 
with $M<1\cdot10^4 M_{\odot}$ in the evolutionary synthesis models to 
estimate ages and masses of stellar groupings.

It is shown, that the method we use for estimating the age 
and mass of the stellar component in star formation regions is applicable 
only for objects, in which the optical radiation from stars coincides with 
the ionized gas emission and for objects without gas emission within the 
area of optical radiation from stars. Note that the number of regions 
with a displacement between the centres of gas emission and photometric 
(stellar) radiation is 3 times greater than the number of regions where 
the optical radiation from stars coincides with the gas emission.

The derived masses of stellar groupings range from $430 M_{\odot}$ in the 
nearby galaxy NGC~628 to $3\cdot10^7 M_{\odot}$ in the distant NGC~7678. 
Most stellar groupings have masses in the range of 
$3\cdot10^3 M_{\odot} - 3\cdot10^5 M_{\odot}$. Two thirds of stellar 
groupings can be attributed to massive star clusters with 
$M>1\cdot10^4 M_{\odot}$.

The range of ages of stellar groupings is from 1 to 560~Myr. One third 
of regions are younger than 10~Myr, and another one-third of objects has 
an age of $10-25$~Myr. The age boundary, at which regions without gas 
emission begin to predominate over objects with H$\alpha$ 
emission, is $15-16$~Myr.

The lower mass estimates for the regions in NGC~628, NGC~3184, and 
NGC~6946 overlaps with the mass interval of the young Milky Way open 
clusters. This is an argument for the existence of a uniform evolutionary 
sequence of extragalactic star formation regions and Galactic open 
clusters at different stages of their evolution.

The introduced $R-{\rm H}\alpha$ index 
$= R+2.5\log F({\rm H}\alpha$+[N\,{\sc ii}]) can be used when 
EW(H$\alpha$) is not available because of absence of spectral 
observations of star formation regions for their analysis.

\section*{Acknowledgments}

We are extremely grateful to the anonymous referee for his/her helpful 
and constructive comments. The authors would like to thank 
A.~E.~Piskunov (Institute of Astronomy of Russian Academy of Sciences) 
for helpful consultions. The authors acknowledge the use of the 
HyperLeda database (\url{http://leda.univ-lyon1.fr}), the NASA/IPAC 
Extragalactic Database (\url{http://ned.ipac.caltech.edu}), Strasbourg 
Astronomical Data Center (CDS, \url{https://cds.u-strasbg.fr}), the 
Sloan Digital Sky Survey (SDSS, \url{http://www.sdss.org}), the Padova 
group online server CMD (\url{http://stev.oapd.inaf.it}), the European 
Southern Observatory Munich Image Data Analysis System ({\sc eso-midas}, 
\url{http://www.eso.org/sci/software/esomidas}), and {\sc SExtractor} 
program (\url{http://sextractor.sourceforge.net}). This study was 
supported by the Russian Foundation for Basic Research (project 
no.~20-02-00080). This research has been supported by the 
Interdisciplinary Scientific and Educational School of Moscow University 
'Fundamental and Applied Space Research'.

\section*{Data availability}
The catalogue is presented as the additional online material to 
this paper in the MNRAS website. It is also available in electronic form 
at \url{http://lnfm1.sai.msu.ru/\~gusev/sfr_cat.html}. Some of the 
images are available in NASA/IPAC Extragalactic Database at 
\url{http://ned.ipac.caltech.edu}. Spectral data are available in 
the Sloan Digital Sky Survey at \url{http://www.sdss.org}, Strasbourg 
Astronomical Data Center at \url{https://cds.u-strasbg.fr}, or in 
corresponding papers. Our own $UBVRI$ and H$\alpha$ observational data 
can be shared on reasonable request to the corresponding author.

\end{document}